# Divergence-Free Adaptive Mesh Refinement for Magnetohydrodynamics

By
**Dinshaw S. Balsara**
Physics Department, University of Notre Dame
(dbalsara@nd.edu)

**Mailing Address:**
Physics Department
College of Science
University of Notre Dame
225 Nieuwland Science Hall
Notre Dame, IN 46556

**Abstract**

Several physical systems such as non-relativistic and relativistic magnetohydrodynamics (MHD), radiation MHD, electromagnetics and incompressible hydrodynamics satisfy Stoke's law type equations for the divergence-free evolution of vector fields. In this paper we present a full-fledged scheme for the second order accurate, divergence-free evolution of vector fields on an adaptive mesh refinement (AMR) hierarchy. We focus here on adaptive mesh MHD. However, the scheme has applicability to the other systems of equations mentioned above. The scheme is based on making a significant advance in the divergence-free reconstruction of vector fields. In that sense, it complements the earlier work of Balsara and Spicer (1999) where we discussed the divergence-free time-update of vector fields which satisfy Stoke's law type evolution equations. Our advance in divergence-free reconstruction of vector fields is such that it reduces to the total variation diminishing (TVD) property for one-dimensional evolution and yet goes beyond it in multiple dimensions. For that reason, it is extremely suitable for the construction of higher order Godunov schemes for MHD. Both the two dimensional and three dimensional reconstruction strategies are developed. A slight extension of the divergence-free reconstruction procedure yields a divergence-free prolongation strategy for prolonging magnetic fields on AMR hierarchies. Divergence-free restriction is also discussed. Because our work is based on an integral formulation, divergence-free restriction and prolongation can be carried out on AMR meshes with any integral refinement ratio, though we specialize the expressions for the most popular situation where the refinement ratio is two. Furthermore, we pay attention to the fact that in order to efficiently evolve the MHD equations on AMR hierarchies, the refined meshes must evolve in time with time steps that are a fraction of their parent mesh's time step. An electric field correction strategy is presented for use on AMR meshes. The electric field correction strategy helps preserve the divergence-free evolution of the magnetic field even when the time steps are sub-cycled on refined meshes. The above-mentioned innovations have been implemented in Balsara's RIEMANN framework for parallel, self-adaptive computational astrophysics which supports both non-relativistic and relativistic MHD. Several rigorous, three dimensional AMR-MHD test problems with strong discontinuities have been run with the RIEMANN framework showing that the strategy works very well. In our AMR-MHD scheme the adaptive mesh hierarchy can change in response to discontinuities that move rapidly with respect to the mesh. Time-step sub-cycling permits efficient processing of the AMR hierarchy. Our AMR-MHD scheme parallelizes very well as shown by Balsara and Norton (Parallel Computing, 2001, vol. 27, pg. 37).

**I) Introduction**

It is well-known that several problems in science and engineering would benefit greatly from using a multi-scale strategy for their solution. Brandt [16] and Berger and Oliger [11] showed the worth of adaptive mesh refinement (AMR) in scientific and engineering calculations. Berger and Colella [12] made a dramatic advance by showing the usefulness of AMR techniques for solving the Euler equations in the presence of strong shocks. Such equations satisfy a conservation law of the form:

$$\frac{\partial \mathbf{U}}{\partial t} + \frac{\partial \mathbf{F}}{\partial x} + \frac{\partial \mathbf{G}}{\partial y} + \frac{\partial \mathbf{H}}{\partial z} = 0 \qquad (1.1)$$

Here **U** is the vector of conserved variables and **F**, **G** and **H** are the flux vectors in the three directions. By using Gauss's law, equations of the form given in eqn. (1.1) can be written in a fully conservative form. The conservative form makes it possible for these systems of equations to admit physically meaningful discontinuous solutions such as shocks and contact discontinuities. For that reason, Colella [20] designed a multidimensional, conservative, higher order Godunov scheme for the solution of such equations. Such higher order Godunov schemes are widely thought to be the most reliable and accurate schemes for solving conservative hyperbolic systems. The loss of strict conservation causes equations of the form given in eqn. (1.1) to develop unphysical solutions. In order to have a conservative solution strategy on an AMR mesh hierarchy Berger and Colella [12] had to make numerous innovations. First, they used a bilinear interpolation strategy for prolonging the solution as well as the boundary information from a coarse level to the child meshes that constitute the fine level. They did, however, mention that by using the very same reconstruction strategy that is used in the underlying higher order Godunov scheme they would have obtained conservative prolongation. Second, they had to design a restriction strategy for transferring the more accurate fine mesh solution to the parent coarse meshes. They did this by using a volume-weighted restriction strategy that was very close in spirit to the volume-averaged representation of variables in the underlying higher order Godunov scheme. Third, they realized that the restriction could cause a loss of conservation at the interfaces between a fine and a coarse mesh. They showed that this loss of conservation could be rectified by using a consistent set of fluxes at the fine-coarse interface. Consistency of the fluxes could be restored via a flux correction step. As a result, exactly conservative evolution of the Euler equations is guaranteed in the strategy of Berger and Colella [12] as long as the levels are properly nested, one within the other. They also realized that Courant number limitations on a fine mesh would cause it to take time steps that are an integral fraction smaller than the coarse mesh time steps. This integral fraction is the reciprocal of the refinement ratio "r". To retain efficiency in processing the entire AMR hierarchy they made a fourth innovation which was to permit the fine mesh's time steps to be sub-cycled so that a fine mesh finds itself time-synchronized with its parent coarse meshes every time it has taken "r" time steps. The AMR innovations by Berger and Colella [12] were very successful because they were based on a small number of well-thought out and easily implemented principles. As a result, AMR strategies for solving conservation laws routinely use the techniques developed in Berger and Colella [12]. Balsara and Norton [8] showed that such AMR schemes ( and their MHD variants) can be easily and efficiently parallelized on parallel supercomputers with large numbers of processors, thereby enhancing the utility of such techniques. The resulting AMR-MHD scheme has been implemented in Balsara's RIEMANN framework for parallel, self-adaptive computational astrophysics.

While several physical systems satisfy a set of equations that can be written in conservation form, not all physical systems satisfy such equations. Several important systems of equations like the Maxwell equations for electromagnetism and the MHD equations satisfy a different update equation given by Faraday's law which has the form:

$$\frac{\partial \mathbf{B}}{\partial t} + c \nabla \times \mathbf{E} = 0 \quad (1.2)$$

where **B** is the magnetic field, **E** is the electric field and c is the speed of light. Other important systems of equations that satisfy an update equation of this form include the equations of incompressible flow, radiation MHD and relativistic MHD. In the specific cases of ideal MHD, radiation MHD and relativistic MHD the electric field is given by

$$\mathbf{E} = -\frac{1}{c} \mathbf{v} \times \mathbf{B} \quad (1.3)$$

where **v** is the fluid velocity. Examination of eqn. (1.2) shows that it is fundamentally different from eqn. (1.1). Unlike eqn. (1.1), eqn. (1.2) does not require the components of the magnetic field to be conserved in a volume-averaged sense. Eqn. (1.2) does predict, via application of Stoke's law, that the magnetic field remains divergence-free. The divergence-free evolution of the magnetic field in eqn. (1.2) has an importance in the design of computational schemes for MHD that is on par with our requirement that systems of equations that satisfy eqn. (1.1) be discretized in a conservative fashion. Loss of divergence-free evolution has been shown to result in unphysical solutions by several authors, for example see Balsara and Spicer [7] and Toth [43]. Brackbill and Barnes [14] and Brackbill [15] have shown that violating the $\nabla \cdot \mathbf{B} = 0$ constraint leads to unphysical plasma transport orthogonal to the magnetic field as well as a loss of momentum and energy conservation. This comes about because violating the constraint results in the addition of extra source terms in the momentum and energy equations. Schemes for numerical MHD that do not satisfy the $\nabla \cdot \mathbf{B} = 0$ constraint and also violate momentum and energy conservation have indeed been designed, see for example the scheme of Powell [37]. In this respect, the demonstration by Toth [43] that the MHD scheme of Powell [37] produces unphysical results is particularly convincing and shows the practical utility of the reasoning in Brackbill and Barnes [14] and Brackbill [15].

Powell et al [38] did attempt to use the scheme of Powell [37] for solving the MHD equations on a three dimensional AMR hierarchy. Their reasonable initial expectation was that as the mesh is refined, the discretization error decreases, resulting in smaller violations of the $\nabla \cdot \mathbf{B} = 0$ constraint on successively refined meshes. However, Powell et al [38] found that violation of the $\nabla \cdot \mathbf{B} = 0$ constraint on one level in the AMR hierarchy causes the $\nabla \cdot \mathbf{B} = 0$ constraint to be violated to the same relative extent on all levels in the AMR hierarchy. Thus using a sequence of finer meshes, with their smaller discretization errors, did not cause the undivided difference of the divergence of the magnetic field to decrease relative to the mean magnetic field on the successively refined meshes. The problems pointed out by Toth [43], therefore, degraded the solution to equal extents on all levels of the AMR hierarchies that were generated by Powell et al [38]. Thus we conclude that mesh refinement is an inadequate foil against a scheme for numerical MHD that is not divergence-free. A violation of divergence-free evolution of the magnetic field on one level in an AMR hierarchy eventually pollutes the solution at all levels of the AMR hierarchy. Physical systems that satisfy equations of the form given in eqn. (1.2) are of great importance in several areas of science and engineering. This has caused us to devote careful attention to the divergence-free evolution of vector fields on an AMR hierarchy. In a fashion that is completely analogous to Berger and Colella [12], the purpose of this paper is to design and catalogue a small number of well-thought out and easily implemented principles for the accurate, efficient and divergence-free time-evolution of vector fields that are governed by an equation of the form given in eqn. (1.2) on AMR hierarchies. We take ideal MHD as our system of interest in the rest of this paper because it is of great importance in astrophysical and space physics applications. For the rest of this paper we will also assume that a refinement ratio of two is used in constructing the AMR hierarchy though the formulation developed here is easily extended to other refinement ratios.

The conservation that is implicit in the structure of eqn. (1.1) is held to be an important enough property for the Euler equations that almost all schemes enforce a discrete version of it at each zone of the computational mesh. The divergence-free evolution of vector fields that is implicit in the structure of eqn. (1.2) is also held to be an equally important property. Thus most practitioners have seen the value of enforcing a discrete version of divergence-free time-evolution of magnetic fields at each zone. This can be done via a "staggered mesh magnetic field transport algorithm" where the magnetic field components are collocated at the face-centers of each zone and the electric field components are collocated at the edge-centers of the zones. Stoke's law is then applied to eqn. (1.2) to yield a discrete time-update strategy for the face-centered magnetic fields. Because it follows from a discrete version of Stoke's law, the resulting discrete time-update strategy clearly shows that if the magnetic field is divergence-free at the beginning of a time step, it will remain so at the end of the time step. Such an algorithm was first proposed by Yee [45] for the transport of electromagnetic fields. Later Brecht et al [17] combined such a discretization strategy with a non-linear Flux Corrected Transport (FCT) based flux limiter in their global MHD modelling of the Earth's magnetosphere. Evans and Hawley [24] then implemented the hybrid scheme developed by Brecht et al [17] in their artificial viscosity based formulation, coining the term "constrained transport". Contemporaneously with Evans and Hawley [24], DeVore [23] applied the same constrained transport scheme to a FCT algorithm. Stone and Norman [42] made a variant of the scheme of Evans and Hawley [24] where they split the MHD eigenstructure into a 2X2 hyperbolic sub-system for the Alfven waves. This makes the entire fluctuation in the magnetic field propagate at the Alfven speed. While some hyperbolic systems can be split into smaller sub-systems, the MHD equations do not permit such a split, see Jeffrey and Taniuti [30]. Thus splitting the MHD eigenstructure into a 2X2 sub-system by Stone and Norman [42] constitutes an algorithmic flaw in their ZEUS code for astrophysical MHD. It results, amongst other things, in post-shock oscillations developing in magnetosonic shocks of even modest strength unless an unacceptably large artificial viscosity is used. On AMR hierarchies a flaw in the underlying solver gets magnified because the solution is propagated at all levels. For that reason we would discourage the use of such solvers for AMR-MHD applications.

While eqn. (1.2) implies divergence-free evolution of magnetic fields, the equations of MHD are indeed hyperbolic equations and can be viewed on a dimension by dimension basis as having a conservative form. This fact has been used by numerous authors to design accurate, robust and reliable higher order Godunov schemes for numerical MHD. A brief list includes Brio and Wu [18], Zachary, Malagoli and Colella [46], Powell [37], Dai and Woodward [21], Ryu and Jones [40], Roe and Balsara [39] and Balsara [1,2]. Balsara [5] has also designed higher order Godunov schemes for relativistic MHD. Almost all of these authors have shown at least some awareness of the fact that it is important to preserve the divergence-free aspect of the magnetic field. Dai and Woodward [22], Ryu et al [41] and Balsara and Spicer [7] showed that simple extensions of higher order Godunov schemes permit one to formulate divergence-free time-update strategies for the magnetic field. Londrillo and Del Zanna [34] have shown that a staggered mesh formulation of the form used by the above three references is fundamental to divergence-free evolution of the magnetic field. Toth [43] has made a comparative study of such schemes and found the scheme of Balsara and Spicer [7] to be one of the most accurate second order schemes that he tested. It is, therefore, our goal to draw on the accuracy and robustness of higher order Godunov schemes and the staggered mesh formulation of Balsara and Spicer [7] to design a robust and accurate divergence-free scheme for AMR-MHD.

Balsara and Spicer [7] formulated a divergence-free time-update strategy for the magnetic field, which we briefly review in Section II while simultaneously extending it to include resistive MHD. However, we did not formulate a strategy for making a divergence-free reconstruction of the magnetic field. Experience with the Euler equations has shown that the reconstruction strategy of the underlying scheme is important for making conservative prolongation of solutions on AMR hierarchies. For that reason, we formulate a divergence-free reconstruction strategy in Section III. In Section IV we show that a slight extension of the divergence-free reconstruction strategy of Section III yields a general-purpose divergence-free prolongation strategy for use on AMR hierarchies. Berger and Colella [12] also designed a volume-weighted restriction strategy for restricting fine mesh solutions to coarse meshes. In Section IV we too design an analogous strategy for area-weighted restriction of magnetic fields from fine meshes to coarse meshes. Berger and Colella [12] also provided a flux correction strategy that needs to be applied at the interfaces between fine and coarse meshes after the fine mesh's time steps have been sub-cycled. Their flux correction strategy was essential for ensuring conservative updates on the AMR hierarchy. In Section V we provide an analogous electric field correction strategy that needs to be applied at the interfaces between fine and coarse meshes after the fine mesh's time steps have been sub-cycled. The electric field correction strategy ensures divergence-free update of magnetic fields on AMR hierarchies. In Section VI we show that our scheme for divergence-free time-evolution of magnetic fields on AMR hierarchies works very well for several stringent MHD test problems even in the presence of strong magnetosonic shocks. In Section VII we offer some conclusions.

**II) Review and Extension of the Numerical MHD Scheme of Balsara and Spicer [7]**

Balsara and Spicer [7] constructed a divergence-free scheme for ideal MHD. In this section we recapitulate the essentials of their scheme while at the same time showing that it can be naturally extended to resistive, non-ideal MHD. The MHD equations can be cast in a conservative form that is suited for the design of higher order Godunov schemes. In that form they become:

$$\frac{\partial \mathbf{U}}{\partial t} + \frac{\partial \mathbf{F}}{\partial x} + \frac{\partial \mathbf{G}}{\partial y} + \frac{\partial \mathbf{H}}{\partial z} + \frac{\partial \mathbf{F}_r}{\partial x} + \frac{\partial \mathbf{G}_r}{\partial y} + \frac{\partial \mathbf{H}_r}{\partial z} = 0 \qquad (2.1)$$

where $\mathbf{F}$, $\mathbf{G}$ and $\mathbf{H}$ are the ideal fluxes and $\mathbf{F}_r$, $\mathbf{G}_r$ and $\mathbf{H}_r$ are the fluxes from the resistive terms. In the resistive case the electric field that is to be used in eqn. (1.2) is given by:

$$\mathbf{E} = -\frac{1}{c} \mathbf{v} \times \mathbf{B} + \frac{c}{4\pi\sigma} \nabla \times \mathbf{B} \qquad (2.2)$$

Written out explicitly, eqn. (2.1) becomes:

$$\frac{\partial}{\partial t}\begin{pmatrix}\rho\\ \rho v_x\\ \rho v_y\\ \rho v_z\\ \mathcal{E}\\ B_x\\ B_y\\ B_z\end{pmatrix} + \frac{\partial}{\partial x}\begin{pmatrix}\rho v_x\\ \rho v_x^2 + P + \mathbf{B}^2/8\pi - B_x^2/4\pi\\ \rho v_x v_y - B_x B_y/4\pi\\ \rho v_x v_z - B_x B_z/4\pi\\ (\mathcal{E}+P+\mathbf{B}^2/8\pi)v_x - B_x(\mathbf{v}\cdot\mathbf{B})/4\pi\\ 0\\ (v_x B_y - v_y B_x)\\ -(v_z B_x - v_x B_z)\end{pmatrix}$$

$$+ \frac{\partial}{\partial y}\begin{pmatrix}\rho v_y\\ \rho v_x v_y - B_x B_y/4\pi\\ \rho v_y^2 + P + \mathbf{B}^2/8\pi - B_y^2/4\pi\\ \rho v_y v_z - B_y B_z/4\pi\\ (\mathcal{E}+P+\mathbf{B}^2/8\pi)v_y - B_y(\mathbf{v}\cdot\mathbf{B})/4\pi\\ -(v_x B_y - v_y B_x)\\ 0\\ (v_y B_z - v_z B_y)\end{pmatrix} + \frac{\partial}{\partial z}\begin{pmatrix}\rho v_z\\ \rho v_x v_z - B_x B_z/4\pi\\ \rho v_y v_z - B_y B_z/4\pi\\ \rho v_z^2 + P + \mathbf{B}^2/8\pi - B_z^2/4\pi\\ (\mathcal{E}+P+\mathbf{B}^2/8\pi)v_z - B_z(\mathbf{v}\cdot\mathbf{B})/4\pi\\ (v_z B_x - v_x B_z)\\ -(v_y B_z - v_z B_y)\\ 0\end{pmatrix}$$

$$+ \frac{\partial}{\partial x}\begin{pmatrix}0\\ 0\\ 0\\ 0\\ 0\\ 0\\ -\frac{c^2}{4\pi\sigma}(\nabla\times B)_z\\ \frac{c^2}{4\pi\sigma}(\nabla\times B)_y\end{pmatrix} + \frac{\partial}{\partial y}\begin{pmatrix}0\\ 0\\ 0\\ 0\\ 0\\ \frac{c^2}{4\pi\sigma}(\nabla\times B)_z\\ 0\\ -\frac{c^2}{4\pi\sigma}(\nabla\times B)_x\end{pmatrix} + \frac{\partial}{\partial z}\begin{pmatrix}0\\ 0\\ 0\\ 0\\ 0\\ -\frac{c^2}{4\pi\sigma}(\nabla\times B)_y\\ \frac{c^2}{4\pi\sigma}(\nabla\times B)_x\\ 0\end{pmatrix} = 0$$

(2.3)

where $\mathcal{E} = \rho v^2 + P/(\gamma-1) + \mathbf{B}^2/8\pi$ is the total energy and $\sigma$ is the conductivity of the plasma. We see that the flux components of the ideal MHD terms as well as the resistive terms obey the following symmetries:

$$F_7 = -G_6, \quad F_8 = -H_6, \quad G_8 = -H_7,$$
$$F_{r,7} = -G_{r,6}, \quad F_{r,8} = -H_{r,6}, \quad G_{r,8} = -H_{r,7}$$

(2.4)

The Balsara and Spicer [7] scheme is based on realizing that there is a dualism between the fluxes that are produced by a higher order Godunov scheme and the electric fields that are needed in eqn. (1.2). The dualism is put to use by capitalizing on the symmetries in eqn. (2.4). In a higher order Godunov scheme that is spatially and temporally second order accurate, the flux variables are available at the center of each zone's face using a straightforward higher order Godunov scheme. These fluxes are spatially second order and temporally centered. The last three components of the **F**, **G** and **H** fluxes can also be reinterpreted as electric fields in our dual approach. The electric fields are needed at the edge centers as shown in Figure 1 and are to be used to update the face-centered magnetic fields. Thus the Godunov fluxes as well as the contributions from the non-ideal terms are directly assigned to the edge centers as follows (eqns. (2.5) to (2.7) should not be viewed as matrix equations) :

$$E^{n+1/2}_{x,\,i,j+1/2,k+1/2} = \frac{1}{4c}\begin{pmatrix} H^{n+1/2}_{7,\,i,j,k+1/2} + H^{n+1/2}_{7,\,i,j+1,k+1/2} \\ - G^{n+1/2}_{8,\,i,j+1/2,k} - G^{n+1/2}_{8,\,i,j+1/2,k+1} \end{pmatrix}$$
$$+ \frac{c}{8\pi\sigma}\left(\frac{1}{\Delta y}\left(B^{n}_{z,\,i,j+1,k+1/2} - B^{n}_{z,\,i,j,k+1/2}\right) - \frac{1}{\Delta z}\left(B^{n}_{y,\,i,j+1/2,k+1} - B^{n}_{y,\,i,j+1/2,k}\right)\right)$$
$$+ \frac{c}{8\pi\sigma}\left(\frac{1}{\Delta y}\left(B^{n+1}_{z,\,i,j+1,k+1/2} - B^{n+1}_{z,\,i,j,k+1/2}\right) - \frac{1}{\Delta z}\left(B^{n+1}_{y,\,i,j+1/2,k+1} - B^{n+1}_{y,\,i,j+1/2,k}\right)\right)$$
(2.5)

$$E^{n+1/2}_{y,\,i+1/2,j,k+1/2} = \frac{1}{4c}\begin{pmatrix} F^{n+1/2}_{8,\,i+1/2,j,k} + F^{n+1/2}_{8,\,i+1/2,j,k+1} \\ - H^{n+1/2}_{6,\,i,j,k+1/2} - H^{n+1/2}_{6,\,i+1,j,k+1/2} \end{pmatrix}$$
$$+ \frac{c}{8\pi\sigma}\left(\frac{1}{\Delta z}\left(B^{n}_{x,\,i+1/2,j,k+1} - B^{n}_{x,\,i+1/2,j,k}\right) - \frac{1}{\Delta x}\left(B^{n}_{z,\,i+1,j,k+1/2} - B^{n}_{z,\,i,j,k+1/2}\right)\right)$$
$$+ \frac{c}{8\pi\sigma}\left(\frac{1}{\Delta z}\left(B^{n+1}_{x,\,i+1/2,j,k+1} - B^{n+1}_{x,\,i+1/2,j,k}\right) - \frac{1}{\Delta x}\left(B^{n+1}_{z,\,i+1,j,k+1/2} - B^{n+1}_{z,\,i,j,k+1/2}\right)\right)$$
(2.6)

$$E^{n+1/2}_{z,\,i+1/2,j+1/2,k} = \frac{1}{4c}\begin{pmatrix} G^{n+1/2}_{6,\,i,j+1/2,k} + G^{n+1/2}_{6,\,i+1,j+1/2,k} \\ - F^{n+1/2}_{7,\,i+1/2,j,k} - F^{n+1/2}_{7,\,i+1/2,j+1,k} \end{pmatrix}$$
$$+ \frac{c}{8\pi\sigma}\left(\frac{1}{\Delta x}\left(B^{n}_{y,\,i+1,j+1/2,k} - B^{n}_{y,\,i,j+1/2,k}\right) - \frac{1}{\Delta y}\left(B^{n}_{x,\,i+1/2,j+1,k} - B^{n}_{x,\,i+1/2,j,k}\right)\right)$$
$$+ \frac{c}{8\pi\sigma}\left(\frac{1}{\Delta x}\left(B^{n+1}_{y,\,i+1,j+1/2,k} - B^{n+1}_{y,\,i,j+1/2,k}\right) - \frac{1}{\Delta y}\left(B^{n+1}_{x,\,i+1/2,j+1,k} - B^{n+1}_{x,\,i+1/2,j,k}\right)\right)$$
(2.7)

In the above equations as well as in Figure 1 we have assumed a Cartesian mesh with edges of size $\Delta x$, $\Delta y$ and $\Delta z$. Balsara and Spicer [7] do treat more general geometries for the ideal MHD case and the results developed here can also be easily extended to general geometries. Notice that in eqns. (2.5) to (2.7) we have discretized the resistive terms so that they have a semi-implicit form. This should make them numerically stable for any choice of time step. As a result the time step can be set by the Courant number that is permitted by the underlying hyperbolic system solver. Notice too that the resistive terms are spatially and temporally second order accurate. The resistive terms have an elliptic form which makes the implicit part in eqns. (2.5) to (2.7) amenable to solution via any iterative multigrid or Krylov sub-space method. The divergence-free restriction and prolongation strategies that we have developed in succeeding sections should be very important in the construction of iterative schemes for divergence-free treatment of resistive MHD. We will take up issues associated with resistive MHD in a subsequent paper. In this paper we stay focussed on the ideal MHD case.

The zone-centered variables, i.e. the first five components of the vector **U** in eqn. (2.1), can be updated in the normal fashion in which conserved variables are routinely updated in a higher order Godunov scheme. The magnetic fields are updated by applying a discrete version of Stoke's law to eqn. (1.2) which yields:

$$B_{x, i+1/2, j, k}^{n+1} = B_{x, i+1/2, j, k}^{n} - \frac{c \Delta t}{\Delta y \Delta z} \left( \begin{array}{c} \Delta z\, E_{z, i+1/2, j+1/2, k}^{n+1/2} - \Delta z\, E_{z, i+1/2, j-1/2, k}^{n+1/2} \\ + \Delta y\, E_{y, i+1/2, j, k-1/2}^{n+1/2} - \Delta y\, E_{y, i+1/2, j, k+1/2}^{n+1/2} \end{array} \right) \quad (2.8)$$

$$B_{y, i, j-1/2, k}^{n+1} = B_{y, i, j-1/2, k}^{n} - \frac{c \Delta t}{\Delta x \Delta z} \left( \begin{array}{c} \Delta x\, E_{x, i, j-1/2, k+1/2}^{n+1/2} - \Delta x\, E_{x, i, j-1/2, k-1/2}^{n+1/2} \\ + \Delta z\, E_{z, i-1/2, j-1/2, k}^{n+1/2} - \Delta z\, E_{z, i+1/2, j-1/2, k}^{n+1/2} \end{array} \right) \quad (2.9)$$

$$B_{z, i, j, k+1/2}^{n+1} = B_{z, i, j, k+1/2}^{n} - \frac{c \Delta t}{\Delta x \Delta y} \left( \begin{array}{c} \Delta x\, E_{x, i, j-1/2, k+1/2}^{n+1/2} - \Delta x\, E_{x, i, j+1/2, k+1/2}^{n+1/2} \\ + \Delta y\, E_{y, i+1/2, j, k+1/2}^{n+1/2} - \Delta y\, E_{y, i-1/2, j, k+1/2}^{n+1/2} \end{array} \right) \quad (2.10)$$

**III) Divergence-Free TVD Reconstruction of Magnetic Fields**

It has long been known from AMR simulations of the Euler equations that conservative TVD reconstruction was one of the key ingredients in the conservative update of flow variables on AMR hierarchies. In like fashion, we suspect that divergence-free TVD reconstruction of vector fields is one of the key ingredients in the divergence-free update of magnetic fields on AMR hierarchies. In this section we analyze the problem of divergence-free TVD reconstruction of magnetic fields for use on AMR hierarchies. Our method is based on the staggered mesh formulation for the divergence-free time-integration of magnetic fields that is given in Balsara and Spicer [7]. Toth [43] has presented a zone-centered formulation for the divergence-free time-integration of magnetic fields which is easily shown to be a mere averaging of the scheme of Balsara and Spicer [7]. We work within the context of the staggered mesh formulation because it has the following advantages over Toth's zone-centered formulation for AMR-MHD work :

1) The control volumes over which the discrete divergence-free condition is satisfied in Toth's formulation cannot exactly cover a logically rectilinear mesh. As pointed out by Evans and Hawley [24], the inflow or outflow of plasma from the boundary of the mesh can be an important source of introducing divergence into the magnetic field. When the control volumes for the conserved variables and the discrete divergence-free condition are coincident, the divergence-free condition is easy to satisfy even at the boundaries. When that is not the case, satisfying the divergence-free condition exactly at the boundaries is not possible. For that reason, Toth's formulation cannot take naturally to general boundary conditions.

2) The issue of control volumes also plays a role in AMR. The reason is that when the grid lines, the lines of all control volumes and the lines that delineate refinement patches in AMR are all aligned with each other one can enforce several constraints, including the divergence-free one, quite trivially. That is not the case when the control volumes on which the divergence-free constraint is enforced are not aligned with the zones of the mesh.

3) One might want to resort to multigrid cycling for resistive MHD or for implicit MHD. In those cases it is very useful to have a mesh hierarchy where the boundaries of the control volumes at coarser levels in the multigrid hierarchy are coincident with the boundaries of the control volumes at finer levels. All the staggered mesh formulations satisfy this property, the zone-centered formulation of Toth does not satisfy this property. Furthermore, the methods built up in the next section permit one to carry out divergence-free restriction and prolongation of magnetic fields on such meshes.

In the first sub-section we provide a detailed discussion of the problem of divergence-free reconstruction of magnetic fields in two dimensions. In the second sub-section we do the same for divergence-free reconstruction of magnetic fields in three dimensions. The problem of divergence-free prolongation of magnetic fields on AMR hierarchies will be discussed in the next section and will be shown to be an extension of the reconstruction theory developed here. Because divergence-free TVD reconstruction represents a significant contribution to the theory of reconstructing divergence-free vector fields, we develop it thoroughly in this section. This thorough development will also help in the next section where we discuss divergence-free prolongation of magnetic fields on AMR hierarchies.

**III.1) Divergence-free Reconstruction of Magnetic Fields in Two Dimensions**

In this sub-section we focus on the divergence-free, spatially second order accurate reconstruction of the magnetic field in two dimensions. The reason for wanting to catalogue such a reconstruction is that it is much simpler to demonstrate the reconstruction procedure with mathematical details in two dimensions than to do the same in three dimensions. The reconstruction strategy that is developed in this sub-section would also be very useful for two dimensional AMR-MHD codes. The three dimensional extension of these ideas follows trivially and we shall give formulae for the three dimensional case in the next sub-section. Most importantly, we wish to demonstrate that there is an equivalence between conventional TVD reconstruction and the reconstruction developed here. We start with the x and y components of the magnetic field which are assumed to be given at the x and y zone faces of a Cartesian mesh. Let $B_x^+ \equiv B_{x,\, i+1/2,\, j}$ denote the magnetic field at the zone's upper x face and let $B_x^- \equiv B_{x,\, i-1/2,\, j}$ denote the magnetic field at the zone's lower x face. Likewise, let $B_y^+ \equiv B_{y, i, j+1/2}$ denote the magnetic field at the zone's upper y face and let $B_y^- \equiv B_{y, i, j-1/2}$ denote the magnetic field at the zone's lower y face. In order to make a second order accurate reconstruction of the magnetic field in the zone of interest, we need to minimally fit piecewise linear profiles for $B_x^\pm$ in the x zone faces. Similarly, we need to fit piecewise linear profiles for $B_y^\pm$ in the y zone faces. Such piecewise linear profiles can be fitted without changing the fact that the divergence of the magnetic field evaluated over the zone faces is zero. These profiles can be obtained via a slope limiter and in the simple case where a minmod limiter is used they are given by

$$\Delta_y B_x^\pm = \mathrm{minmod}\left( B_{x, i\pm1/2, j+1} - B_{x, i\pm1/2, j} , B_{x, i\pm1/2, j} - B_{x, i\pm1/2, j-1} \right) \tag{3.1}$$

$$\Delta_x B_y^\pm = \mathrm{minmod}\left( B_{y, i+1, j\pm1/2} - B_{y, i, j\pm1/2} , B_{y, i, j\pm1/2} - B_{y, i-1, j\pm1/2} \right) \tag{3.2}$$

Different limiters can be used in place of the minmod limiter and we use the minmod limiter only as an instantiation. One can even use the piecewise-linear WENO interpolation from Jiang and Shu [31] to provide a form of non-oscillatory reconstruction which reduces the clipping at extrema. It will be shown later that there are certain distinct advantages to be gained by fitting limited linear profiles instead of unlimited linear profiles. We locate the origin at the zone-center of the zone being considered. As a result, the variation of $B_x$ and $B_y$ in the upper and lower x and y zone faces is given by:

$$B_x \left( x = \pm \Delta x/2 , y \right) = B_x^\pm + \frac{\Delta_y B_x^\pm}{\Delta y} y \tag{3.3}$$

$$B_y \left( x , y = \pm \Delta y/2 \right) = B_y^\pm + \frac{\Delta_x B_y^\pm}{\Delta x} x \tag{3.4}$$

To simplify the notation, we set the origin at the zone center. The zone has edges of size $\Delta x$ and $\Delta y$.

The problem of reconstructing the magnetic field components in the zone given by $[-\Delta x/2 , \Delta x/2] \times [-\Delta y/2 , \Delta y/2]$ reduces to fitting functions on the interior of the zone so that they match the four linear profiles for $B_x$ and $B_y$ given above in the x and y zone faces. It is not acceptable to fit any two arbitrary functions for $B_x$ and $B_y$. This is because the divergence-free aspect for the reconstructed fields will not be retained in the interior of the zone in that case. Instead we wish to fit two polynomial functions, one for $B_x$ and one for $B_y$, which have the special property that their divergence is exactly zero at all points within the zone. Should we be successful in finding such polynomials, we will be able to evaluate transverse magnetic field components on either side of each zone face. This would be very useful for providing the left and right states for a Riemann solver. We can also use such polynomials to make area-weighted prolongation of the magnetic field components to any of the faces of a refined zone that may lie within the zone of interest. Such a prolongation of the magnetic field will be inherently divergence-free because it is derived from a reconstruction that is divergence-free in a continuous sense. It is easy to see that using piecewise linear polynomials for the two functions would prove inadequate. Thus we consider a quadratic representation which we write as:

$$B_x(x, y) = a_0 + a_x x + a_y y + a_{xx} x^2 + a_{xy} xy + a_{yy} y^2 \qquad (3.5)$$

$$B_y(x, y) = b_0 + b_x x + b_y y + b_{xx} x^2 + b_{xy} xy + b_{yy} y^2 \qquad (3.6)$$

Because we have to fit linear profiles at the zone faces we set

$$a_{yy} = b_{xx} = 0 \qquad (3.7)$$

Imposing a divergence-free condition in a continuous sense gives three further constraints on the coefficients of the two polynomials given above. The constraints are:

$$a_x + b_y = 0 \; ; \; 2 a_{xx} + b_{xy} = 0 \; ; \; a_{xy} + 2 b_{yy} = 0 \qquad (3.8)$$

When the reconstructed magnetic fields in eqns. (3.5) and (3.6) satisfy the constraints given in eqns. (3.7) and (3.8), the reconstructed magnetic fields will be divergence-free in the entire zone.

After accounting for all the constraints in eqns. (3.7) and (3.8) we see that the two polynomials for $B_x$ and $B_y$ that are given by eqns. (3.5) and (3.6) have seven independent coefficients. The variation of $B_x$ and $B_y$ in the x and y zone faces is given by eqns. (3.3) and (3.4) and can be described by exactly seven independent numbers since the divergence-free aspect of the magnetic field requires an integral constraint to be satisfied in discrete form. That constraint is given by:

$$\left(B_x^+ - B_x^-\right) \Delta y + \left(B_y^+ - B_y^-\right) \Delta x = 0 \qquad (3.9)$$

The reconstruction polynomials for $B_x$ and $B_y$ that are given by eqns. (3.5) and (3.6) should match the linear profiles that have been fitted to the boundaries in eqns. (3.3) and (3.4). Making this match at the boundaries is tantamount to equating linear combinations of the seven independent coefficients in eqns. (3.5) and (3.6) to the seven independent numbers in eqns. (3.3) and (3.4) and yields seven independent equations. The seven equations form a linear system which can be inverted to yield the seven independent coefficients in eqns. (3.5) and (3.6). On carrying out the algebra we get:

$$a_x = -b_y = \frac{\left(B_x^+ - B_x^-\right)}{\Delta x} = -\frac{\left(B_y^+ - B_y^-\right)}{\Delta y} \qquad (3.10)$$

$$a_y = \frac{1}{2}\left(\frac{\Delta_y B_x^+}{\Delta y} + \frac{\Delta_y B_x^-}{\Delta y}\right) \tag{3.11}$$

$$b_x = \frac{1}{2}\left(\frac{\Delta_x B_y^+}{\Delta x} + \frac{\Delta_x B_y^-}{\Delta x}\right) \tag{3.12}$$

$$a_{xy} = -2\, b_{yy} = \frac{1}{\Delta x}\left(\frac{\Delta_y B_x^+}{\Delta y} - \frac{\Delta_y B_x^-}{\Delta y}\right) \tag{3.13}$$

$$b_{xy} = -2\, a_{xx} = \frac{1}{\Delta y}\left(\frac{\Delta_x B_y^+}{\Delta x} - \frac{\Delta_x B_y^-}{\Delta x}\right) \tag{3.14}$$

$$a_0 = \frac{\left(B_x^+ + B_x^-\right)}{2} - a_{xx}\frac{\Delta x^2}{4} \tag{3.15}$$

$$b_0 = \frac{\left(B_y^+ + B_y^-\right)}{2} - b_{yy}\frac{\Delta y^2}{4} \tag{3.16}$$

Thus we have demonstrated that we can indeed make an exact, divergence-free reconstruction of the magnetic field inside the zone that matches the linear profiles of $B_x$ and $B_y$ at the zone's faces. This reconstruction can now be used in an area-weighted sense to prolong the magnetic field to the faces of any refined zone that lies within the zone being considered. Realize too that if the same reconstruction strategy is applied to two zones that share a face then it will produce the same linear profile for the magnetic field component on the shared face. As a result, the refined zone to which we want to prolong magnetic field components can even have faces that coincide with coarse zone faces. We have, therefore, shown that in two dimensions it is possible to make a divergence-free prolongation of the magnetic field to any refined zone. For the two dimensional case, the reconstruction theory developed in this sub-section can be directly used for divergence-free prolongation on AMR hierarchies and no further extension of the theory is needed.

When all the variations are restricted to one dimension our divergence-free reconstruction strategy reduces to standard TVD reconstruction. This will be explicitly demonstrated for the three dimensional case in the next sub-section. An analogous demonstration can also be shown to hold true for the two dimensional case.

**III.2) Divergence-free Reconstruction of Magnetic Fields in Three Dimensions**

To establish some notation we refer the reader to Figure 1. As in the previous sub-section we establish a short-form notation that is given by

$$B_x^\pm \equiv B_{x,\, i\pm 1/2,\, j,\, k}\ ;\ B_y^\pm \equiv B_{y,\, i,\, j\pm 1/2,\, k}\ ;\ B_z^\pm \equiv B_{z,\, i,\, j,\, k\pm 1/2} \tag{3.17}$$

The limited slopes for these variables are given by

$$\Delta_y B_x^\pm = \text{minmod}\left(B_{x, i\pm1/2, j+1, k} - B_{x, i\pm1/2, j, k}, B_{x, i\pm1/2, j, k} - B_{x, i\pm1/2, j-1, k}\right) \quad (3.18)$$

$$\Delta_z B_x^\pm = \text{minmod}\left(B_{x, i\pm1/2, j, k+1} - B_{x, i\pm1/2, j, k}, B_{x, i\pm1/2, j, k} - B_{x, i\pm1/2, j, k-1}\right) \quad (3.19)$$

$$\Delta_x B_y^\pm = \text{minmod}\left(B_{y, i+1, j\pm1/2, k} - B_{y, i, j\pm1/2, k}, B_{y, i, j\pm1/2, k} - B_{y, i-1, j\pm1/2, k}\right) \quad (3.20)$$

$$\Delta_z B_y^\pm = \text{minmod}\left(B_{y, i, j\pm1/2, k+1} - B_{y, i, j\pm1/2, k}, B_{y, i, j\pm1/2, k} - B_{y, i, j\pm1/2, k-1}\right) \quad (3.21)$$

$$\Delta_x B_z^\pm = \text{minmod}\left(B_{z, i+1, j, k\pm1/2} - B_{z, i, j, k\pm1/2}, B_{z, i, j, k\pm1/2} - B_{z, i-1, j, k\pm1/2}\right) \quad (3.22)$$

$$\Delta_y B_z^\pm = \text{minmod}\left(B_{z, i, j+1, k\pm1/2} - B_{z, i, j, k\pm1/2}, B_{z, i, j, k\pm1/2} - B_{z, i, j-1, k\pm1/2}\right) \quad (3.23)$$

In a fashion that is entirely analogous to the two dimensional case, the piecewise linear variation of $B_x$, $B_y$ and $B_z$ in the x, y and z zone faces are given by:

$$B_x\left(x = \pm\Delta x/2, y, z\right) = B_x^\pm + \frac{\Delta_y B_x^\pm}{\Delta y} y + \frac{\Delta_z B_x^\pm}{\Delta z} z \quad (3.24)$$

$$B_y\left(x, y = \pm\Delta y/2, z\right) = B_y^\pm + \frac{\Delta_x B_y^\pm}{\Delta x} x + \frac{\Delta_z B_y^\pm}{\Delta z} z \quad (3.25)$$

$$B_z\left(x, y, z = \pm\Delta z/2\right) = B_z^\pm + \frac{\Delta_x B_z^\pm}{\Delta x} x + \frac{\Delta_y B_z^\pm}{\Delta y} y \quad (3.26)$$

To simplify the notation, we set the origin at the zone center. The zone has edges of size $\Delta x$, $\Delta y$, and $\Delta z$.

The reconstructed fields in the interior of the zone given by $\left[-\Delta x/2, \Delta x/2\right] \times \left[-\Delta y/2, \Delta y/2\right] \times \left[-\Delta z/2, \Delta z/2\right]$ which match the linear variation of the fields on the zone faces can be written as :

$$B_x(x, y, z) = a_0 + a_x x + a_y y + a_z z + a_{xx} x^2 + a_{xy} x y + a_{xz} x z \quad (3.27)$$

$$B_y(x, y, z) = b_0 + b_x x + b_y y + b_z z + b_{xy} x y + b_{yy} y^2 + b_{yz} y z \quad (3.28)$$

$$B_z(x, y, z) = c_0 + c_x x + c_y y + c_z z + c_{xz} x z + c_{yz} y z + c_{zz} z^2 \quad (3.29)$$

Imposing the divergence-free condition in a continuous sense gives four further constraints on the coefficients of the three polynomials given by eqns. (3.27) to (3.29). The constraints are:

$$a_x + b_y + c_z = 0 \; ; \; 2a_{xx} + b_{xy} + c_{xz} = 0 \; ; \; a_{xy} + 2b_{yy} + c_{yz} = 0 \; ;$$
$$a_{xz} + b_{yz} + 2c_{zz} = 0 \tag{3.30}$$

After accounting for all the constraints in eqn. (3.30) we see that the above polynomials for $B_x$, $B_y$ and $B_z$ that are given by eqns. (3.27) to (3.29) have seventeen independent coefficients. We insist on matching one field component along with its two transverse variations on each of the six faces of the zone, see eqns. (3.24) to (3.26). This yields eighteen conditions out of which one is dependent on the others, being given by the discrete divergence-free condition:

$$\left(B_x^+ - B_x^-\right)\Delta y\, \Delta z + \left(B_y^+ - B_y^-\right)\Delta x\, \Delta z + \left(B_z^+ - B_z^-\right)\Delta x\, \Delta y = 0 \tag{3.31}$$

This results in seventeen conditions that need to be satisfied by the reconstructed fields which, we recall, have seventeen independent coefficients. This reduces the problem to straightforward algebraic manipulation. While inverting a 17X17 matrix may seem like a daunting task, a little examination of the matrix shows that it splits up into several independent 2X2 sub-systems, each of which is easily inverted. On carrying out the algebra we get:

$$a_x = \frac{\left(B_x^+ - B_x^-\right)}{\Delta x} \tag{3.32}$$

$$a_y = \frac{1}{2}\left(\frac{\Delta_y B_x^+}{\Delta y} + \frac{\Delta_y B_x^-}{\Delta y}\right) \tag{3.33}$$

$$a_z = \frac{1}{2}\left(\frac{\Delta_z B_x^+}{\Delta z} + \frac{\Delta_z B_x^-}{\Delta z}\right) \tag{3.34}$$

$$a_{xy} = \frac{1}{\Delta x}\left(\frac{\Delta_y B_x^+}{\Delta y} - \frac{\Delta_y B_x^-}{\Delta y}\right) \tag{3.35}$$

$$a_{xz} = \frac{1}{\Delta x}\left(\frac{\Delta_z B_x^+}{\Delta z} - \frac{\Delta_z B_x^-}{\Delta z}\right) \tag{3.36}$$

$$a_{xx} = -\frac{1}{2}\left(b_{xy} + c_{xz}\right) \tag{3.37}$$

$$a_0 = \frac{\left(B_x^+ + B_x^-\right)}{2} - a_{xx}\frac{\Delta x^2}{4} \tag{3.38}$$

To obtain the formulae for the coefficients in eqn. (3.28), make the following replacements a → b, b → c, c → a, x → y, y → z and z → x in eqns. (3.32) to (3.38) above. Similarly, to obtain the formulae for the coefficients in eqn. (3.29), make the replacements a → c, b → a, c → b, x → z, y → x and z → y in eqns. (3.32) to (3.38) above. Allow self-evident permutations of the form $c_{zx} \equiv c_{xz}$.

The volume-averaged magnetic field component in the x-direction in the zone being considered is given by

$$\langle B_x \rangle_{vol-avg} = \frac{\left(B_x^+ + B_x^-\right)}{2} - a_{xx}\frac{\Delta x^2}{6} \tag{3.39}$$

We see that eqn. (3.39) is not exactly the mean x-component of the field evaluated at the zone center. The inclusion of the divergence-free reconstruction can cause small changes because of the need to maintain the continuous divergence-free condition within the zone. However, it is also noteworthy that the difference is $O\left(\Delta x^2\right)$. Thus the zone-averaged fields evaluated above are comparable to the mean fields up to second order accuracy. Furthermore, the coefficient $a_{xx}$ is based on using a limiting procedure and should, therefore, stay bounded. Previous practitioners had always used the mean fields evaluated at the zone center even in situations where they might have wanted to use the volume-averaged magnetic field variables. We now see, post facto, that their choice was a good one as long as they were designing schemes that were second order accurate.

It is very useful to ask what our reconstruction strategy produces when all the variations are restricted to one dimension? Thus, say for example, that we choose that direction to be the x direction. In that case $B_x$ becomes a constant ( because of the divergence-free condition) and $B_y$ and $B_z$ can have variations along the x direction provided we have $B_y^+ = B_y^-$ and $B_z^+ = B_z^-$. Then many of the polynomial coefficients become zero. The only non-zero coefficients are given by

$$a_0 = B_x^+ = B_x^- = \text{constant} \; ; \; b_0 = B_y^+ = B_y^- \; ; \; c_0 = B_z^+ = B_z^- \; ;$$
$$b_x = \frac{\Delta_x B_y^+}{\Delta x} \; ; \; c_x = \frac{\Delta_x B_z^+}{\Delta x} \tag{3.40}$$

The expressions for $b_x$ and $c_x$ in the equation (3.40) make it very clear that it is useful to use limited slopes for $\Delta_x B_y^+$ and $\Delta_x B_z^+$. When we use limited slopes, our reconstruction strategy reduces to MUSCL interpolation for one dimensional variations, see vanLeer [44]. The fluxes that will be produced when using this divergence-free reconstruction strategy will, therefore, be exactly the same as the fluxes that will be produced by a zone-centered, higher order Godunov scheme for numerical MHD in the limiting case of one dimensional variations. This clearly shows the equivalence of our new scheme for numerical MHD, which uses our new three-dimensional divergence-free reconstruction strategy, to a standard, zone-centered, higher order Godunov scheme that uses standard TVD reconstruction when all the variations are restricted to lie along one direction. Our demonstration also serves to strengthen the claim made in Balsara and Spicer [7] that in numerical MHD there is an equivalence between certain components of the higher order Godunov fluxes and the electric field.

It is worthwhile to make several interesting points about the divergence-free reconstruction strategy that has been developed in the previous and present sub-sections:

1) vanLeer [44] showed that the solution of a hyperbolic system in conservative form could be reduced to three simple steps :- i) make TVD reconstruction, ii) apply Riemann solver and iii) make conservative update. The innovations that have been made in this section permit us to formulate a similar three step plan for hyperbolic systems with divergence-free vector fields that follow a Stokes law type evolution equation :- i) Thus we can use the results of this section to make a divergence-free TVD reconstruction of the vector fields. ii) Riemann solvers for many of these systems of equations have already been designed and are catalogued in point 7) below. They can be applied to obtain upwinded fluxes. iii) Divergence-free update strategies for vector fields have also been designed in Balsara and Spicer [7] and in Section II of the present paper. They can be used to make a divergence-free update of the vector field.

2) Our divergence-free reconstruction strategy for magnetic fields, along with any TVD or WENO reconstruction of the zone-centered fluid dynamical variables, can be used in conjunction with a temporally second order accurate Runge Kutta or predictor-corrector time-stepping strategy to yield a spatially and temporally second order accurate scheme for MHD. A thorough examination of such schemes and their properties will be carried out in subsequent work.

3) Our reconstruction strategy has been developed in the simple case where the limiter is applied directly to the magnetic field components. This is tantamount to applying the limiter to the primitives, with the magnetic field components being taken as the primitives along with the usual fluid dynamical primitives of density, pressure and velocities. Harten [27] showed the value of using the characteristic variables in the limiter. The slopes used in our scheme can also be obtained from such a characteristic reconstruction strategy, albeit at a greater computational cost.

4) Earlier in this section we have shown that the divergence-free reconstruction reduces to standard TVD reconstruction when all the variations are restricted to lie along one direction. A further connection can be made with dimension by dimension TVD reconstruction by focussing on the linear terms in eqns. (3.27) to (3.29). Eqns. (3.33) and (3.34) clearly show that the linear slopes are limited, thereby establishing the connection. Furthermore, eqn. (3.32) and similar expressions for $b_y$ and $c_z$ have to assume the form that they do in order for the divergence-free condition in eqn. (3.31) to be satisfied. The quadratic terms in eqns. (3.27) to (3.29) are comprised of differences of limited quantities and can, therefore, be expected to stay bounded.

5) vanLeer [44] also carried out the stability analysis for the advection of a scalar field when the limiter is applied to the scalar field. VanLeer's stability analysis helped in providing a firm conceptual foundation for zone-centered higher order Godunov schemes. The present work makes it possible to carry out the stability analysis for the divergence-free advection of a vector field. This is especially true since the induction equation for MHD is linear in the magnetic field and the velocity. Such a stability analysis will be carried out in subsequent work and will provide a firm conceptual foundation for Stokes-law type update equations.

6) It is possible to take an alternative view of our divergence-free reconstruction strategy. In that view, realize that specifying the seven (in 2d) or seventeen (in 3d) independent coefficients of the polynomials is equivalent to specifying the divergence-free field components and their slopes at the zone faces. This then suggests that a Galerkin formulation for MHD would be especially beneficial where one would evolve not just the magnetic field component at each zone face but also its first ( or higher) moments in the plane of that face. This could be done using the discontinuous Galerkin formulations in Cockburn and Shu [19] or Lowrie, Roe and vanLeer [35] and making suitable extensions to those formulations by using the ideas developed here. We will develop this direction in subsequent work.

7) There are similar problems with preserving the divergence of the magnetic field in numerical electrodynamics, see Yee [45]; incompressible flow, see Harlow and Welch [26]; relativistic MHD, see Balsara [5] and radiation MHD, see Balsara [3,4]. Hence those application areas would also benefit from the strategies developed here.

**IV) Divergence-Free Restriction and Prolongation on AMR Hierarchies**

The problem of divergence-free restriction of magnetic fields in a staggered mesh formulation is very simple. It consists of making an area weighted average of the magnetic field component that is collocated on the faces of the fine mesh and assigning it to the corresponding face of the coarse mesh. Such a step should be carried out ( along with the electric field correction step that is detailed in the next section) whenever the fine and coarse meshes are temporally synchronized. If the field on the fine mesh is divergence-free, this combination of area-weighted restriction and electric field correction will result in a divergence-free field on the coarse mesh. The previous statement will be proved in the next section.

The problem of divergence-free prolongation is far more intricate and interesting than the restriction problem. To see the intricacy, see Figure 2 which shows a fine mesh that abuts a coarse mesh. The dotted lines indicate the region where we want to lay down a new fine mesh. As one can see, a part of the new fine mesh will have zone faces that are aligned with the zone faces of the old fine mesh. The magnetic field components on those faces of the new fine mesh can be assigned by straight injection of the magnetic field components from the old fine mesh. This will be a divergence-free assignment. The layer of coarse mesh zones that do not share a face with the fine mesh can also make a divergence-free prolongation to the new fine mesh. This can be done by first fitting piecewise linear slopes in the transverse directions to the field components on each of the faces of those zones. We can then use the reconstruction theory developed in the previous section to reconstruct the divergence-free magnetic field in each of the coarse zones of interest. This reconstruction can then be used to make a divergence-free prolongation to the new fine mesh. One would naively think that the same strategy can be applied to the coarse mesh zones that share a face with the fine mesh. However, such a naive application of the same strategy to these zones would produce a problem. To see the source of the problem, consider the x-faces that are shared between the old fine mesh and the coarse mesh in Figure 2. Figure 3 provides a blow-up of such an interface. Fitting a linear variation in the y and z directions to the coarse mesh's $B_x$ component on those shared faces only provides three degrees of freedom on the coarse mesh face. However, the coarse mesh's x-faces overlap with four x-faces on the old fine mesh. It is impossible for the piecewise-linear profile on the coarse mesh's face to match those four values in an integral sense. To achieve such a match we need to include a "yz" variation in the coarse mesh's $B_x$ component. Thus when a coarse mesh's face does not coincide with four fine mesh faces, we fit to it a profile given by eqns. (3.24) to (3.26) using the piecewise-linear profiles given by eqns. (3.18) to (3.23). However, in the most general case, the coarse mesh's faces can coincide with four fine mesh faces. In that case matching with the old fine mesh's four face-centered fields requires us to use profiles of the form :

$$B_x \left( x = \pm \Delta x/2 , y, z \right) = B_x^\pm + \frac{\Delta_y B_x^\pm}{\Delta y} y + \frac{\Delta_z B_x^\pm}{\Delta z} z + \frac{\Delta_{yz} B_x^\pm}{\Delta y \, \Delta z} y \, z \qquad (4.1)$$

$$B_y \left( x , y = \pm \Delta y/2, z \right) = B_y^\pm + \frac{\Delta_x B_y^\pm}{\Delta x} x + \frac{\Delta_z B_y^\pm}{\Delta z} z + \frac{\Delta_{xz} B_y^\pm}{\Delta x \, \Delta z} x \, z \qquad (4.2)$$

$$B_z \left( x , y, z = \pm \Delta z/2 \right) = B_z^\pm + \frac{\Delta_x B_z^\pm}{\Delta x} x + \frac{\Delta_y B_z^\pm}{\Delta y} y + \frac{\Delta_{xy} B_z^\pm}{\Delta x \, \Delta y} x \, y \qquad (4.3)$$

To simplify the notation, we set the origin at the center of the coarse zone being considered. The coarse zone has edges of size $\Delta x$, $\Delta y$, and $\Delta z$. Our strategy for fitting profiles to the coarse mesh's zone faces can now be summarized as follows : For coarse mesh zone faces that do not overlie old fine mesh zone faces, we simply set the cross terms (i.e. the last terms) in eqns. (4.1) to (4.3) to zero and make piecewise-linear profiles using eqns. (3.18) to (3.23). On the other hand, for coarse mesh zone faces that do overlie old fine mesh zone faces, we do not make linear profiles but rather use the more general profiles given in eqns. (4.1) to (4.3) with non-zero cross terms. The values of the coefficients are then set by requiring that the variation on the coarse mesh's zone face matches the four field components on the underlying fine mesh faces. An example will help. Say that the $x = -\Delta x/2$ face in eqn. (4.1) overlies four fine mesh faces. In a shorthand notation, let the $B_x$ components on those faces be denoted by:

$$b^-_{x,+,+} = B_x\left(x = -\Delta x/2, y = \Delta y/4, z = \Delta z/4\right);$$
$$b^-_{x,+,-} = B_x\left(x = -\Delta x/2, y = \Delta y/4, z = -\Delta z/4\right);$$
$$b^-_{x,-,+} = B_x\left(x = -\Delta x/2, y = -\Delta y/4, z = \Delta z/4\right);$$
$$b^-_{x,-,-} = B_x\left(x = -\Delta x/2, y = -\Delta y/4, z = -\Delta z/4\right)$$

(4.4)

Then the coefficients in eqn. (4.1) are given by:

$$B^-_x = \frac{1}{4}\left(b^-_{x,+,+} + b^-_{x,-,-} + b^-_{x,+,-} + b^-_{x,-,+}\right)$$
$$\Delta_{yz}B^-_x = 4\left(b^-_{x,+,+} + b^-_{x,-,-} - b^-_{x,+,-} - b^-_{x,-,+}\right)$$
$$\Delta_y B^-_x = \left(b^-_{x,+,+} - b^-_{x,-,-} + b^-_{x,+,-} - b^-_{x,-,+}\right)$$
$$\Delta_z B^-_x = \left(b^-_{x,+,+} - b^-_{x,-,-} - b^-_{x,+,-} + b^-_{x,-,+}\right)$$

(4.5)

On comparing eqns. (4.1) and (3.24) we see that we now have two extra terms in $B_x\left(x = -\Delta x/2, y, z\right)$ and $B_x\left(x = \Delta x/2, y, z\right)$. Both those terms have a "yz" variation in the x-faces. Matching the two "yz" variations at the two coarse zone boundaries requires that we add two extra terms in the reconstruction polynomial for $B_x(x, y, z)$ in eqn. (3.27). The minimal terms that give us that extra variation are given by adding $a_{yz} y z + a_{xyz} x y z$ to the polynomial expression for $B_x(x, y, z)$. However, to have a divergence-free reconstruction in the coarse zone we have to do two more things:
1) Add a term $b_{yyz} y^2 z$ in the polynomial expression for $B_y(x, y, z)$ in eqn. (3.28).
2) Add a term $c_{yzz} y z^2$ in the polynomial expression for $B_z(x, y, z)$ in eqn. (3.29).

In an analogous fashion, on comparing eqns. (4.2) and (3.25) we see that we now have two extra terms in $B_y\left(x, y = -\Delta y/2, z\right)$ and $B_y\left(x, y = \Delta y/2, z\right)$. Both those terms have a "xz" variation in the y-faces. To match that variation we again do three things:
1) Add a term $b_{xz} x z + b_{xyz} x y z$ in the polynomial expression for $B_y(x, y, z)$ in eqn. (3.28).
2) Add a term $a_{xxz} x^2 z$ in the polynomial expression for $B_x(x, y, z)$ in eqn. (3.27).
3) Add a term $c_{xzz} x z^2$ in the polynomial expression for $B_z(x, y, z)$ in eqn. (3.29).

In a similar fashion, on comparing eqns. (4.3) and (3.26) we see that we now have two extra terms in $B_z\left(x, y, z = -\Delta z/2\right)$ and $B_z\left(x, y, z = \Delta z/2\right)$. Both those terms have a "xy" variation in the z-faces. To match that variation we again do three things:
1) Add a term $c_{xy} x y + c_{xyz} x y z$ in the polynomial expression for $B_z(x, y, z)$ in eqn. (3.29).
2) Add a term $a_{xxy} x^2 y$ in the polynomial expression for $B_x(x, y, z)$ in eqn. (3.27).
3) Add a term $b_{xyy} x y^2$ in the polynomial expression for $B_y(x, y, z)$ in eqn. (3.28).

Putting all the terms together in eqns. (3.27) to (3.29) we get polynomials that are well-suited for prolongation as follows :

$$B_x(x, y, z) = a_0 + a_x x + a_y y + a_z z + a_{xx} x^2 + a_{xy} x y + a_{xz} x z \\ + a_{yz} y z + a_{xyz} x y z + a_{xxz} x^2 z + a_{xxy} x^2 y \quad (4.6)$$

$$B_y(x, y, z) = b_0 + b_x x + b_y y + b_z z + b_{xy} x y + b_{yy} y^2 + b_{yz} y z \\ + b_{xz} x z + b_{yyz} y^2 z + b_{xyz} x y z + b_{xyy} x y^2 \quad (4.7)$$

$$B_z(x, y, z) = c_0 + c_x x + c_y y + c_z z + c_{xz} x z + c_{yz} y z + c_{zz} z^2 \\ + c_{xy} x y + c_{yzz} y z^2 + c_{xzz} x z^2 + c_{xyz} x y z \quad (4.8)$$

Along with eqn. (3.30), the divergence-free condition now yields the following extra constraints:

$$a_{xyz} + 2 b_{yyz} + 2 c_{yzz} = 0 \quad (4.9)$$

$$2 a_{xxz} + b_{xyz} + 2 c_{xzz} = 0 \quad (4.10)$$

$$2 a_{xxy} + 2 b_{xyy} + c_{xyz} = 0 \quad (4.11)$$

It is important to realize that the "yz" variations in the two x-faces, the "xz" variations in the two y-faces and the "xy" variations in the two z-faces provide six extra conditions. We have introduced twelve extra polynomial coefficients in eqns. (4.6) to (4.8) which are required to satisfy three constraints. Thus nine of the extra polynomial coefficients are independent. However, we only have six extra conditions at the zone faces with which to fix up nine independent polynomial coefficients. Thus the system of equations is under-determined and we can make simplifying choices for three of the polynomial coefficients. These choices are made so that the resulting equations have as elegant and simple a structure as is possible. We also want the resulting prolongation polynomials to be closest in their structure to the reconstruction polynomials in the previous section. In particular, when the "yz", "xz" and "xy" variations on the x, y and z faces respectively are set to zero, we want our prolongation polynomials to reduce transparently to the reconstruction polynomials in the previous section. These goals can be simply achieved by setting :

$$b_{yyz} = c_{yzz} = - a_{xyz} / 4 \quad (4.12)$$

$$a_{xxz} = c_{xzz} = - b_{xyz} / 4 \quad (4.13)$$

$$a_{xxy} = b_{xyy} = - c_{xyz} / 4 \quad (4.14)$$

Notice that when $a_{xyz} = b_{xyz} = c_{xyz} = 0$ the additional polynomial coefficients trivially become zero. Thus with the above-mentioned simple choice, the equations for the polynomial coefficients become well-determined and the prolongation polynomials also reduce to the appropriate limits.

The polynomial coefficients can now be trivially evaluated. In fact, many of them are still given by expressions that were built up in the previous section, hence we do not repeat them here. Here we simply provide closed form expressions for new polynomial coefficients and coefficients that have changed from their analogues in sub-section III.3.

$$a_y = \frac{1}{2}\left(\frac{\Delta_y B_x^+}{\Delta y} + \frac{\Delta_y B_x^-}{\Delta y}\right) + c_{xyz}\frac{\Delta x^2}{16} \tag{4.15}$$

$$a_z = \frac{1}{2}\left(\frac{\Delta_z B_x^+}{\Delta z} + \frac{\Delta_z B_x^-}{\Delta z}\right) + b_{xyz}\frac{\Delta x^2}{16} \tag{4.16}$$

$$a_{yz} = \frac{1}{2}\left(\frac{\Delta_{yz} B_x^+}{\Delta y\, \Delta z} + \frac{\Delta_{yz} B_x^-}{\Delta y\, \Delta z}\right) \tag{4.17}$$

$$a_{xyz} = \frac{1}{\Delta x}\left(\frac{\Delta_{yz} B_x^+}{\Delta y\, \Delta z} - \frac{\Delta_{yz} B_x^-}{\Delta y\, \Delta z}\right) \tag{4.18}$$

To derive formulae for $b_z$, $b_x$, $b_{xz}$ and $b_{xyz}$ from the formulae for $a_y$, $a_z$, $a_{yz}$ and $a_{xyz}$ respectively, set a → b, b → c, c → a, x → y, y → z, z → x and $\Delta_{yz} \rightarrow \Delta_{xz}$ in the right hand sides of eqns. (4.15) to (4.18) above. Similarly, to derive formulae for $c_x$, $c_y$, $c_{xy}$ and $c_{xyz}$ from the formulae for $a_y$, $a_z$, $a_{yz}$ and $a_{xyz}$ respectively, set a → c, b → a, c → b, x → z, y → x, z → y and $\Delta_{yz} \rightarrow \Delta_{xy}$ in the right hand sides of eqns. (4.15) to (4.18) above.

When the prolongation polynomial for $B_x$ is prolonged in an area-weighted fashion to a face that has $x = x_0$ and y and z extents given by $[y_0, y_1]\times[z_0, z_1]$, the resulting area-averaged value of $B_x$ is given by

$$\begin{aligned}\langle B_x\rangle_{area-avg} &= \left(a_0 + a_x x_0 + a_{xx} x_0^2\right) + \left(a_y + a_{xy} x_0 + a_{xxy} x_0^2\right)\frac{(y_1 + y_0)}{2} \\ &+ \left(a_z + a_{xz} x_0 + a_{xxz} x_0^2\right)\frac{(z_1 + z_0)}{2} \\ &+ \left(a_{yz} + a_{xyz} x_0\right)\frac{(y_1 + y_0)}{2}\frac{(z_1 + z_0)}{2}\end{aligned} \tag{4.19}$$

Similar area-weighted prolongation can be carried out for the y and z components of the field. The previous formula and others like it for the other two directions can be used to make divergence-free prolongation of magnetic field components to the zone faces of refined zones.

It is worthwhile to make several interesting points about the prolongation strategy that we have developed in the present section:

1) Our prolongation strategy is a minimal extension of our reconstruction strategy. When the cross terms on the zone faces are zero in eqns. (4.1) to (4.3) it reduces exactly to the reconstruction scheme in the previous section.

2) The prolongation strategy developed here should be used for initializing the magnetic field on newly formed fine meshes. The reconstruction strategy developed in Section III also has its uses. When a fine level in an AMR hierarchy takes two time steps for every one coarse level time step, we also need to provide the fine level's halo zones with spatially and temporally second order accurate values from the coarse level. Those values can also be provided by using the reconstruction strategy developed in Section III. The reconstruction strategy developed in Section III has the advantage that one does not need to code up the extra logic associated with deciding whether to include or exclude the last terms in eqns. (4.1) to (4.3). This saves some float point operations.

3) Experience with AMR techniques applied to the Euler equations has shown that TVD reconstruction of the conserved variables should be used in order to make conservative prolongation of the fluid variables. We find that our minimal extension of the divergence-free TVD reconstruction strategy yields a divergence-free prolongation strategy for the magnetic fields.

4) Just as Berger and Colella [12] used a volume-weighted averaging for their restriction strategy, we use an area-weighted averaging as our restriction strategy for the magnetic fields. This is because our magnetic fields are fundamentally area-weighted averages just as the conserved variables in a higher order Godunov scheme for the Euler equations are fundamentally volume-weighted averages.

5) Our prolongation strategy is based on an integral formulation and is not dependent on any special algebraic tricks. As a result, it can be easily extended for making divergence-free prolongation on an AMR mesh with any refinement ratio. For example, refinement ratios of four or eight can be achieved by applying the present strategy recursively. For other refinement ratios, new formulas would have to be derived to account for the additional degrees of freedom needed to match the increased number of existing fine-grid faces.

6) The formulae can also be easily extended to any orthogonal mesh geometry, such as cylindrical or spherical, and are not restricted to Cartesian geometry.

7) Peyrard and Villedieu [36] have presented a scheme that works on two dimensional triangulated meshes and preserves the divergence of the magnetic field only in a weak sense. Our prolongation strategy produces a divergence-free reconstruction of the magnetic field at all points in the zone and keeps $\nabla \cdot \mathbf{B} = 0$ up to machine precision. We have also been able to show that the strategy presented here also goes over for triangular and tetrahedral meshes. As a result it is especially useful for unstructured meshes and cut-cell approaches because one can develop exact integrals for fine zone faces that are not perfectly aligned with the coarse zone faces.

8) In two dimensions, say a coarse mesh face overlies two fine mesh faces. In that case, fitting a field value and a piecewise-linear slope in the transverse direction on that coarse mesh face will allow us to match the two field components on the fine mesh faces. Thus the extensions to three dimensional reconstruction that are described in this section are not needed. Therefore, in two dimensions, the reconstruction strategy described in sub-section III.1 also doubles up as a prolongation strategy.

9) The methods developed in this section and the previous one are also important for multigrid processing of the MHD equations on adaptive meshes. Such multigrid processing could turn out to be very important in designing time-implicit schemes for MHD. The work of Jameson [28] and Gropp et al [25] has already demonstrated the importance of multilevel processing for the Euler equations. The work of Balsara [6] has already demonstrated the importance of multigrid methods for the equations of radiative transfer. Thus the importance of multigrid processing for the MHD equations seems almost evident. Multigrid methods would also be important for resistive MHD.

10) Our divergence-free prolongation strategy is actually a divergence-preserving strategy when eqns. (3.32) and similar expressions for $b_y$ and $c_z$ are used without imposing the constraint $a_x + b_y + c_z = 0$. This can be very useful for mildly compressible hydrodynamical problems or incompressible flow problems where wall effects can cause vorticity to be built up on a mesh.

11) Janhunen [29] presents an interesting alternative to the scheme of Powell [37] which tries to satisfy the $\nabla \cdot \mathbf{B} = 0$ constraint up to discretization error, but not machine error. It may be argued that such a scheme might be adequate for AMR-MHD calculations, perhaps without the use of the divergence-free prolongation formulae presented here. Such a scheme would have two sources of magnetic field divergence : a) the underlying scheme itself and b) the prolongation scheme at the interfaces between fine and coarse meshes. The problem with such a scheme is that the divergence of the magnetic field, once generated in whatever fashion, will flow with the plasma. In space physics problems such a divergence of the field builds up at the stagnation point of the magnetosphere and pollutes the solution there. However, the stagnation point is the very place where one is likely to want an accurate solution. Similarly, in accretion problems in astrophysics, the divergence of the field will build up at the central object and pollute the solution there. This is usually the very place where one wants an accurate solution.

**V) The Electric Field Correction Step at Fine-Coarse Interfaces**

In the previous section we briefly discussed the problem of restriction in a staggered mesh formulation. This ensures that a coarse mesh that shares a face with four fine mesh faces will always have the best possible estimate for the magnetic field when the fine and coarse meshes are temporally synchronized. For example, take the coarse mesh in Figure 3 as having edges of size $\Delta x$, $\Delta y$, and $\Delta z$ and take the fine mesh in Figure 3 as having edges that are half that size. The coarse mesh takes a time step of size $\Delta t$ to go from time $t^n$ to time $t^{n+1}$. The time step equation for $B_{1,x}^n$ can be written as

$$B_{1,x}^{n+1} = B_{1,x}^n - \frac{\Delta t}{\Delta y} \left( E_{2,z}^{n+1/2} - E_{1,z}^{n+1/2} \right) - \frac{\Delta t}{\Delta z} \left( E_{1,y}^{n+1/2} - E_{2,y}^{n+1/2} \right) \quad (5.1)$$

The coarse mesh takes two time steps, each of size $\Delta t/2$, to traverse the same time interval. The two time step equations for $b_{1,x}^n$ can be written as

$$b_{1,x}^{n+1/2} = b_{1,x}^n - \frac{\Delta t}{\Delta y} \left( e_{2,z}^{n+1/4} - e_{1,z}^{n+1/4} \right) - \frac{\Delta t}{\Delta z} \left( e_{1,y}^{n+1/4} - e_{3,y}^{n+1/4} \right) \quad (5.2)$$

$$b_{1,x}^{n+1} = b_{1,x}^{n+1/2} - \frac{\Delta t}{\Delta y} \left( e_{2,z}^{n+3/4} - e_{1,z}^{n+3/4} \right) - \frac{\Delta t}{\Delta z} \left( e_{1,y}^{n+3/4} - e_{3,y}^{n+3/4} \right) \quad (5.3)$$

Similar equations can be written for $b_{2,x}^n$, $b_{3,x}^n$, and $b_{4,x}^n$. At a time $t^{n+1}$ the restriction step requires us to synchronize $B_{1,x}^{n+1}$ with $b_{1,x}^{n+1}$, $b_{2,x}^{n+1}$, $b_{3,x}^{n+1}$, and $b_{4,x}^{n+1}$ as follows:

$$B_{1,x}^{n+1} = \frac{1}{4} \left( b_{1,x}^{n+1} + b_{2,x}^{n+1} + b_{3,x}^{n+1} + b_{4,x}^{n+1} \right) \quad (5.4)$$

However, a look at Figure 3 will readily show that restriction, taken by itself, cannot ensure consistency or divergence-free time-evolution. The reason is that $B_{1,x}^{n+1}$ after eqn. (5.4) is applied to it is different from $B_{1,x}^{n+1}$ that was produced by eqn. (5.1). Thus the divergence in the first coarse mesh zone that abuts the fine mesh zones will be non-zero at time $t^{n+1}$ after the restriction step, even if it were zero before the restriction step was applied at time $t^{n+1}$ and at all times prior to $t^{n+1}$, i.e. say at time $t^n$. Examining eqns. (5.2) and (5.3) and others like them shows us that imposing eqn. (5.4) has implicitly caused us to change the update step for $B_{1,x}^n$ from eqn. (5.1) to the following update step:

$$B_{1,x}^{n+1} = B_{1,x}^{n} - \frac{\Delta t}{\Delta y} \left[ \begin{array}{c} \frac{1}{4} \left( e_{3,z}^{n+1/4} + e_{6,z}^{n+1/4} + e_{3,z}^{n+3/4} + e_{6,z}^{n+3/4} \right) \\ - \frac{1}{4} \left( e_{1,z}^{n+1/4} + e_{4,z}^{n+1/4} + e_{1,z}^{n+3/4} + e_{4,z}^{n+3/4} \right) \end{array} \right]$$
$$- \frac{\Delta t}{\Delta z} \left[ \begin{array}{c} \frac{1}{4} \left( e_{1,y}^{n+1/4} + e_{2,y}^{n+1/4} + e_{1,y}^{n+3/4} + e_{2,y}^{n+3/4} \right) \\ - \frac{1}{4} \left( e_{5,y}^{n+1/4} + e_{6,y}^{n+1/4} + e_{5,y}^{n+3/4} + e_{6,y}^{n+3/4} \right) \end{array} \right]$$
(5.5)

where we have assumed that an equation like eqn. (5.4) also holds at time $t^n$, implying that a similar restriction step was enforced at time $t^n$. This is equivalent to replacing the coarse mesh's electric fields with the best spatially and temporally interpolated electric fields from the fine mesh. The loss of divergence-free magnetic field structure on the AMR hierarchy after restriction stems from examining the time-evolution of magnetic field components like $B_{1,y}^n$, $B_{2,y}^n$, $B_{1,z}^n$ and $B_{2,z}^n$ which lie on coarse mesh faces that share an edge with the fine mesh. In the course of temporal evolution from time $t^n$ to time $t^{n+1}$ these magnetic field components have not used the spatially and temporally interpolated electric fields from the fine mesh in their time step equation. From eqn. (5.5) we see that $B_{1,x}^n$ has, on the other hand, used the spatially and temporally interpolated electric fields from the fine mesh in its time step equation. Had we consistently used the same electric fields for updating $B_{1,x}^n$ as well as $B_{1,y}^n$, $B_{2,y}^n$, $B_{1,z}^n$ and $B_{2,z}^n$, the problem would not have arisen and the magnetic field on the coarse mesh would have remained divergence-free after the restriction step. Thus the problem is entirely analogous to the flux correction problem that arises in the AMR processing of the Euler equations as explained in Berger and Colella [12]. The solution is, therefore, also very similar. The solution simply consists of modifying the updated fields $B_{1,y}^{n+1}$, $B_{2,y}^{n+1}$, $B_{1,z}^{n+1}$ and $B_{2,z}^{n+1}$ at time $t^{n+1}$ so that their update is equivalent to our having used the best spatially and temporally interpolated electric fields from the fine mesh. This is done as follows:

$$B_{1,y}^{n+1} = B_{1,y}^{n+1} - \frac{\Delta t}{\Delta x} \left( \frac{1}{4} \left( e_{1,z}^{n+1/4} + e_{4,z}^{n+1/4} + e_{1,z}^{n+3/4} + e_{4,z}^{n+3/4} \right) - E_{1,z}^{n+1/2} \right) \tag{5.6}$$

$$B_{1,z}^{n+1} = B_{1,z}^{n+1} + \frac{\Delta t}{\Delta x} \left( \frac{1}{4} \left( e_{1,y}^{n+1/4} + e_{2,y}^{n+1/4} + e_{1,y}^{n+3/4} + e_{2,y}^{n+3/4} \right) - E_{1,y}^{n+1/2} \right) \tag{5.7}$$

Similar equations can be written for $B_{2,y}^{n+1}$ and $B_{2,z}^{n+1}$.

We call eqns. (5.6) and (5.7) the electric field correction step because they are carried out in a fashion that is analogous to the flux correction step in Berger and Colella [12]. Application of such a step to all the magnetic field components on the coarse mesh that need such a correction step restores the divergence-free magnetic field structure on the coarse mesh after a restriction step. Just as the flux correction step in Berger and Colella [12] restores the consistency of the fluxes at the interface between fine and coarse meshes, the electric field correction step described here restores the consistency of the electric fields at the interface between fine and coarse meshes. In restoring this consistency, the flux correction step ensures conservative evolution of the flow in an AMR hierarchy and, in a completely analogous fashion, the electric field correction step ensures divergence-free evolution of the magnetic fields in an AMR hierarchy. Restoring the consistency of the fluxes and electric fields at the interface between fine and coarse meshes also helps stabilize the time-update strategy for the AMR hierarchy thereby enabling it to evolve with the maximal Courant number that is permitted by the underlying higher order Godunov scheme.

**VI) Results**

The results presented in this section were carried out with the RIEMANN framework for computational astrophysics which incorporates the following advances in parallel, self-adaptive numerical MHD:

1) A fast TVD scheme which draws on the ideas in Roe and Balsara [39], Balsara [1,2] and Balsara and Spicer [7] was used as the underlying hyperbolic system solver. The scheme is not dimensionally swept but rather uses a multistage, second-order accurate Runge-Kutta update strategy which permits a maximal Courant number of 0.4. That Courant number was used on all levels of the AMR hierarchy in all the problems presented here showing that our time step sub-cycling strategy works efficiently. The Balsara and Spicer [7] scheme for divergence-free time-update of magnetic fields was used on each level in the AMR hierarchy.

2) The magnetic fields on each newly formed fine level were initialized in two stages. First, if it was possible to use the magnetic fields from the old fine level, then those fields were assigned to the corresponding locations of the newly formed fine level. Second, for those face-centered magnetic field locations that could not be initialized via the first step, divergence-free prolongation from Section IV was used to prolong the magnetic fields from the parent level to the newly formed fine level. This two stage strategy ensures that each newly formed fine level would always have the best possible representation of magnetic fields that could be assigned to it.

3) Divergence-free reconstruction from Section III was used to provide a divergence-free prolongation of the coarse level magnetic fields to the halo (ghost) zones of the fine level. Spatially and temporally second order accurate, divergence-free interpolation was used so that the halo zones of each fine level always had the best possible representation of the magnetic field that the coarse level could provide even when its time step was sub-cycled. The vanLeer limiter was used in place of the minmod limiter in eqns. (3.18) to (3.23) to permit a slightly smoother structure in the magnetic fields being prolonged.

4) When prolonging the zone-centered, conserved variables we used a TVD reconstruction for each zone from which the conserved variables were to be prolonged. This ensured that we were using a fully conservative prolongation strategy for the conserved variables.

5) Restriction of magnetic field variables was done via the area-weighted restriction strategy given in Section IV. This gives us a divergence-free restriction strategy for the magnetic fields belonging to coarse level zones that overlap fine level zones.

6) Restriction of conserved variables was done via the volume-weighted restriction strategy detailed in Berger and Colella [12]. This gives us a conservative restriction strategy for the conserved variables belonging to coarse level zones that overlap fine level zones.

7) The electric field correction strategy from Section V was used to restore divergence-free magnetic fields to the coarse level zone faces that share a zone edge with the fine-coarse interface.

8) The flux correction strategy from Berger and Colella [12] was used to restore full conservation to the coarse level zones that share a zone face with the fine-coarse interface.

9) As explained in Section V, the electric field and flux correction steps restore consistency to the time-update of the AMR hierarchy. As a result, each mesh in the AMR hierarchy can evolve with the full CFL number that is permitted by the underlying hyperbolic system solver.

10) Flagging strategies described in Lohner [31,32] and Powell et al [38] were used to flag locations in the flow that develop singularities (such as shocks, contact discontinuities and rotational discontinuities) and are, therefore, in need of refinement. The RIEMANN framework actually incorporates a slew of physics-based flagging strategies that are catalogued in Balsara and Norton [8]. The appropriate flagging strategies are automatically swapped in or swapped out based on the physics of the problem being solved.

11) The regridding strategy in Berger and Rigoutsos [13] and Bell et al [10] was used. We modified the regridding strategy slightly to ensure that the levels remain properly nested, one within the other. This is essential for enforcing the electric field and flux correction steps.

12) The parallelization strategy from Balsara and Norton [8] was used to ensure high levels of parallel processing efficiency.

Two test problems, both of which are three dimensional, are presented which illustrate divergence-free AMR-MHD. The first test problem consists of a strong magnetosonic shock interacting with a dense blob of plasma, i.e. a cloud. The second test problem is drawn from supernova remnant (SNR) simulations and consists of setting off a strong explosion in a magnetized medium. A test problem should be simple enough that it can be set up and run quite easily. In that spirit, both test problems have been designed so that they can be set up on rather small root grids and use only a couple of levels of refinement. Just using two levels of refinement in these test problems, however, produces a dramatic improvement in the solutions. There is no restriction on increasing the number of levels of refinement used, at the expense of increased computational time. We have used two levels of refinement here because they permit all the steps in our divergence-free AMR-MHD strategy to be fully exercised.

**VI.a) Shock-Cloud Interaction**

This test problem was motivated by a somewhat similar test problem in Dai and Woodward [22]. A 96X64X64 zone root grid was used to cover a Cartesian domain of extent [-0.5, 1.0]X[-0.5,0.5]X[-0.5,0.5]. The majority of the computational domain was filled with a low density plasma with $\left(\rho, P, v_x, v_y, v_z, B_x, B_y, B_z\right)$ given by $\left(1, 1, 0, 0, 0, 2, 2, 2\right)$. A cloud was initialized at the origin with a radius of 0.3 length units and a density that is ten times larger than that of the low density plasma that initially fills most of the volume. The density of the cloud was given a linear taper of size 1/128 length units. Such a linear taper can only be represented on the second level of refinement in an AMR hierarchy. Except for the density, the remaining flow variables in the cloud matched the variables in its ambient plasma. A Mach 10 magnetosonic shock was initialized at x = -0.45. The post-shock region was initialized with flow variables $\left(\rho, P, v_x, v_y, v_z, B_x, B_y, B_z\right)$ given by $\left(3.91109, 253.724, 13.8388, -0.005, -0.005, 2.0, 7.84323, 7.84323\right)$. The x = -0.5 boundary was treated as an inflow boundary. The x = 1.0 boundary was treated as an outflow boundary. The remaining boundaries were extrapolative so that flow features could propagate off the computational domain if they had outwardly propagating characteristic fields. The problem was run to a time of 0.0653 time units. Two levels of refinement were permitted. The calculation was carried out on a parallel supercomputer using double precision arithmetic. While the RIEMANN framework has achieved significant speedups over the baseline performance numbers reported in Balsara and Norton [8], the scalability is similar to that reported in the previous reference.

Figure 4 shows various flow quantities in the xz plane at a time of 0.0251 units. Figure 4a shows the logarithm of the density which clearly illustrates that the shock has engulfed half the surface area of the cloud. In doing so, a bow shock has been set up in the region downstream of the shock. Because of the cloud's higher density a magnetosonic shock has begun to slowly propagate through and compress the leftward-facing part of the cloud. Figure 4b shows the logarithm of the pressure. Figure 4c shows the Mach number and Figure 4d shows the magnitude of the magnetic field. They all show signatures of the shock structures mentioned above. Figure 4e is a color coded representation of the two levels of refinement used in this test problem. The root grid is colored blue; the first level of refinement is colored yellow; the second level of refinement is colored red. We see that the adaptively refined meshes have tracked the time-evolving magnetosonic shocks. Thus the flagging and regridding routines have constructed the refined meshes in just those regions where the presence of shocks or contact discontinuities would call for them. We also see that the mesh refinement has caused all the discontinuities to be represented very crisply on the computational domain so that the simulation genuinely looks like it was done on a 384X256X256 zone mesh. Figure 4f shows the ratio of the undivided divergence of the magnetic field to the magnitude of the magnetic field. It is a good measure of the build up of divergence in this problem. We see that this ratio is less than $10^{-14}$ which is exactly what we would expect in a calculation that was done in double precision. Figure 4f shows a measure of the undivided divergence in the xz plane. However, in the course of carrying out this three dimensional simulation we monitored the divergence of the magnetic field in the whole volume and found that it has the same order of magnitude as that shown in Figure 4f. We, therefore, verify that our divergence-free AMR-MHD scheme has produced genuinely divergence-free evolution of the magnetic field even in this rather stringent problem that has strong, time-evolving magnetosonic shocks.

Figure 5 shows the same flow quantities that were shown in Figure 4 at a time of 0.0653 units. This is a late time in the evolution of the shock-cloud problem. We see from Figures 5a to 5d that the shock that was propagating through the cloud in Figure 4 has by this late time compressed the whole cloud and passed beyond it. (When that shock reached the rightward-facing boundary of the cloud, it impulsively accelerated the lighter ambient material, forming a jet.) In Figures 5a and 5c we can see the jet connecting the cloud's rear surface to the shock that has passed beyond it. Strong magnetic field structures have formed at the leftward-facing surface of the cloud because of compressional effects. Strong magnetic field structures have also formed behind the cloud in and around the jetted region as a result of a combination of compression and field stretching. The sheared interface at the jet's surface launches shear Alfven waves which show up prominently in Figure 5d. The leftward-facing bow shock becomes weaker as time progresses so that it no longer triggers refinement at these late times. The cloud's boundary, the jet that propagates from the cloud's surface and the strong magnetosonic shock that has completely engulfed the cloud are the three prominent flow structures that clearly trigger refinement as shown in Figure 5e. Figure 5f clearly shows that even at this late time the divergence has not built up substantially. The ratio of the undivided divergence to the magnitude of the magnetic field is still well below $10^{-13}$ ! We have also monitored the divergence of the magnetic field in the whole volume and found that it has the same order of magnitude as that shown in Figure 5f. As a result, we have verified that our AMR-MHD scheme is truly divergence-free over long integration times.

Figures 4f and 5f show a speckled structure for the divergence of the magnetic field and the speckled pattern develops smaller scale structure on the finer meshes. Thus one might have thought that this provides a small wavelength instability which propagates across refinement levels via the restriction and prolongation operators used here. One might even raise the concern that this might provoke a numerical instability. However, we see no evidence for such instabilities. We believe that this is because we have used divergence-free reconstruction of magnetic fields with suitable limiters for prolonging the magnetic field variables.

**VI.b) Supernova Blast Wave**

Our second test problem is motivated by the work of Balsara, Benjamin and Cox [9]. A 64X64X64 zone root grid was used to cover a Cartesian domain of extent [-20, 20]X[-20,20]X[-20,20]. The majority of the domain was initialized with a plasma with flow variables $\left(\rho, P, v_x, v_y, v_z, B_x, B_y, B_z\right)$ given by $\left(1, 0.813428, 0, 0, 0, 3.29006, 0, 0\right)$. An initially spherical supernova blast wave was initialized so that it was centered at the origin and had a radius of 1.25 length units. This radius corresponds to the size of two zones on the root grid. Had such a problem been run on the root grid alone, the resulting supernova blast would have been highly anisotropic because of mesh imprinting. Because the problem was recursively initialized on two further levels of refinement, sufficiently many zones were available on the finest level to produce a simulation that was free of strong mesh imprinting. The plasma within the supernova was initialized with a density of 18.01738355 and a pressure of 1033422.167. The fluid within the sphere that was initially occupied by the supernova was given an outward-going velocity profile that was radially oriented and linearly increasing with the radius. The fluid's speed at a radius of 1.25 length units was 350.582016 units. The boundary conditions play no role in this problem and can be set as periodic or continuitive. The problem was stopped a little before the supernova blast hits the boundary. This corresponds to a stopping time of 0.11 units. Two levels of refinement were permitted in this test problem. The RIEMANN framework was again used in double precision mode for this simulation.

Figure 6 shows various flow quantities in the xz plane at a time of 0.02097 units. Figure 6a shows the logarithm of the density which clearly shows an outward-going fast magnetosonic shock, a contact discontinuity and an inward-going fast magnetosonic shock. Figure 6b shows the logarithm of the pressure which tracks the two shocks quite well. Figure 6c shows the Mach number and Figure 6d shows the magnitude of the magnetic field. The field gets strongly compressed by the explosion, producing the barrel-like structure visible in Figure 6d. Figures 6a to 6d all show signatures of the shock structures mentioned above. Figure 6e is a color coded representation of the two levels of refinement used in this test problem. The color coding is similar to Figure 4e. We see that the refined meshes have tracked the time-evolving magnetosonic shocks. Figure 6f shows the undivided divergence. In this problem we do not plot out the ratio of the undivided divergence to the magnetic field magnitude because the expanding blast wave sweeps out almost all of the magnetic field that was initially present in the central region. We see from Figure 6d, however, that the compressed magnetic field has a magnitude of about 10 units. Thus Figure 6f does confirm that our AMR-MHD simulation of a supernova blast wave has been divergence-free up to machine accuracy. Figure 7 shows the same flow variables that were shown in Figure 6 at a late time of 0.11 time units. The divergence-free evolution of the magnetic field is again confirmed. As in the previous test problem we monitored the evolution of the divergence of the magnetic field in the whole volume and confirmed that the divergence was of the same magnitude as shown in figures 6f and 7f. All the shock structures in Figures 6 and 7 are crisply resolved so that the simulation looks as if it had been done on a 256X256X256 zone mesh. In Balsara, Benjamin and Cox [9] we have actually carried out such simulations on uniform meshes that were much finer than the root grid used here. All the flow features in the simulation reported here were also seen to occur in the high resolution uniform mesh simulations. This gives us further confidence that our RIEMANN framework for AMR-MHD is working well.

The present test problem is an even more stringent test problem for AMR-MHD than the previous one. It is perhaps amongst the most stringent test problems that can be designed in computational astrophysics. We have, therefore, verified that our AMR-MHD scheme, as implemented in the RIEMANN framework for computational astrophysics, is truly divergence-free for some of the most stringent test problems that we have been able to design.

## VII) Conclusions

Several advances in divergence-free AMR-MHD are reported here. We list them below:

1) A general strategy is presented for the time-update of systems of equations that satisfy a Stoke's law type equation on AMR hierarchies. Several such physically important systems are identified with special focus on MHD.

2) Just as Berger and Colella [12] reduced the conservative time-update of the Euler equations on an AMR hierarchy to the application of a few simple steps, we have reduced the divergence-free time-update of the MHD equations on an AMR hierarchy to the application of a few simple steps. The steps have been summarized in Section VI.

3) It is shown that the divergence-free time-update scheme of Balsara and Spicer [7] can be elegantly extended to the resistive MHD case. Various arguments are presented to show that the face-centered collocation of magnetic field components is the collocation of choice for MHD and other systems of equations that satisfy a Stoke's law type time-evolution equation.

4) A significant advance has been made in the divergence-free reconstruction of vector fields. It has been shown that for one-dimensional variations, such a reconstruction strategy reduces to the familiar TVD property. Yet in multiple dimensions, it goes well beyond it.

5) Divergence-free prolongation of magnetic fields on an AMR hierarchy can be carried out via a very slight extension of the divergence-free reconstruction scheme mentioned in the previous point.

6) A divergence-free restriction strategy is presented which consists of using area-weighted averaging of magnetic fields from fine mesh faces.

7) An electric field correction strategy is presented which restores the consistency of electric fields at a fine-coarse interface in the AMR hierarchy.

8) Because of the above four points, the time-step can be sub-cycled on finer meshes without loss of the divergence-free property of the magnetic fields. Points 4) and 5) are also essential for the time-step sub-cycling because they provide for a robust and stable numerical strategy. Another pleasant consequence of the above four points is that each mesh in the AMR hierarchy evolves with the full Courant number that is permitted by the underlying hyperbolic system solver. This ensures that there is no undue loss of efficiency in the time-update of the MHD equations on AMR hierarchies.

9) The above-mentioned innovations have been incorporated in the RIEMANN framework for parallel, self-adaptive computational astrophysics. Several stringent test problems have been presented and it is shown that the method presented in this paper for AMR-MHD is truly divergence-free.

**Acknowledgements**

The author gratefully acknowledges discussions with D. Spicer and P. MacNeice. The use of computer time at NCSA is also acknowledged. Support from the NSF's ACR grant # aci-9982160 is also gratefully acknowledged.

**Figure Captions**

Figure 1 shows the collocation of the magnetic fields at the control volume's faces and the collocation of electric fields at the control volume's edges.

Figure 2 shows an old fine mesh (shown with red lines) that abuts a coarse mesh (shown with blue lines). The location of the newly introduced fine mesh (shown with dashed green lines) is such that it overlaps the old fine and coarse meshes. Straight injection can be used to assign magnetic field values from the old fine mesh to the new fine mesh. The layer of zones for which the reconstruction strategy from Section III can be used to assign magnetic field values to the new fine mesh is also shown. The coarse mesh zones that abut fine mesh faces need the prolongation strategy that is developed in Section IV to assign magnetic field values to the new fine mesh.

Figure 3 shows the interface between a fine and a coarse mesh. Face-centered magnetic fields that are needed in the restriction step are shown. Edge-centered electric fields that are needed for the electric field correction step are also shown.

Figure 4 shows selected variables in the central x-z plane from the shock-cloud test problem at a rather early time of 0.0251. Figure 4a shows $\log_{10}$ (density). Figure 4b shows $\log_{10}$ (pressure). Figure 4c shows the Mach number. Figure 4d shows the magnitude of the magnetic field. Figure 4e shows a color-coded representation of the levels in the AMR hierarchy. Figure 4f shows the ratio of the undivided divergence of the magnetic field to the total magnetic field.

Figure 5 shows selected variables in the central x-z plane from the shock-cloud test problem at a rather late time of 0.0653. Figure 5a shows $\log_{10}$ (density). Figure 5b shows $\log_{10}$ (pressure). Figure 5c shows the Mach number. Figure 5d shows the magnitude of the magnetic field. Figure 5e shows a color-coded representation of the levels in the AMR hierarchy. Figure 5f shows the ratio of the undivided divergence of the magnetic field to the total magnetic field.

Figure 6 shows selected variables in the central x-z plane from the Supernova blast test problem at an early time of 0.02097. Figure 6a shows $\log_{10}$ (density). Figure 6b shows $\log_{10}$ (pressure). Figure 6c shows the Mach number. Figure 6d shows the magnitude of the magnetic field. Figure 6e shows a color-coded representation of the levels in the AMR hierarchy. Figure 6f shows the undivided divergence of the magnetic field.

Figure 7 shows selected variables in the central x-z plane from the Supernova blast test problem at a late time of 0.11. Figure 7a shows $\log_{10}$ (density). Figure 7b shows $\log_{10}$ (pressure). Figure 7c shows the Mach number. Figure 7d shows the magnitude of the magnetic field. Figure 7e shows a color-coded representation of the levels in the AMR hierarchy. Figure 7f shows the undivided divergence of the magnetic field.

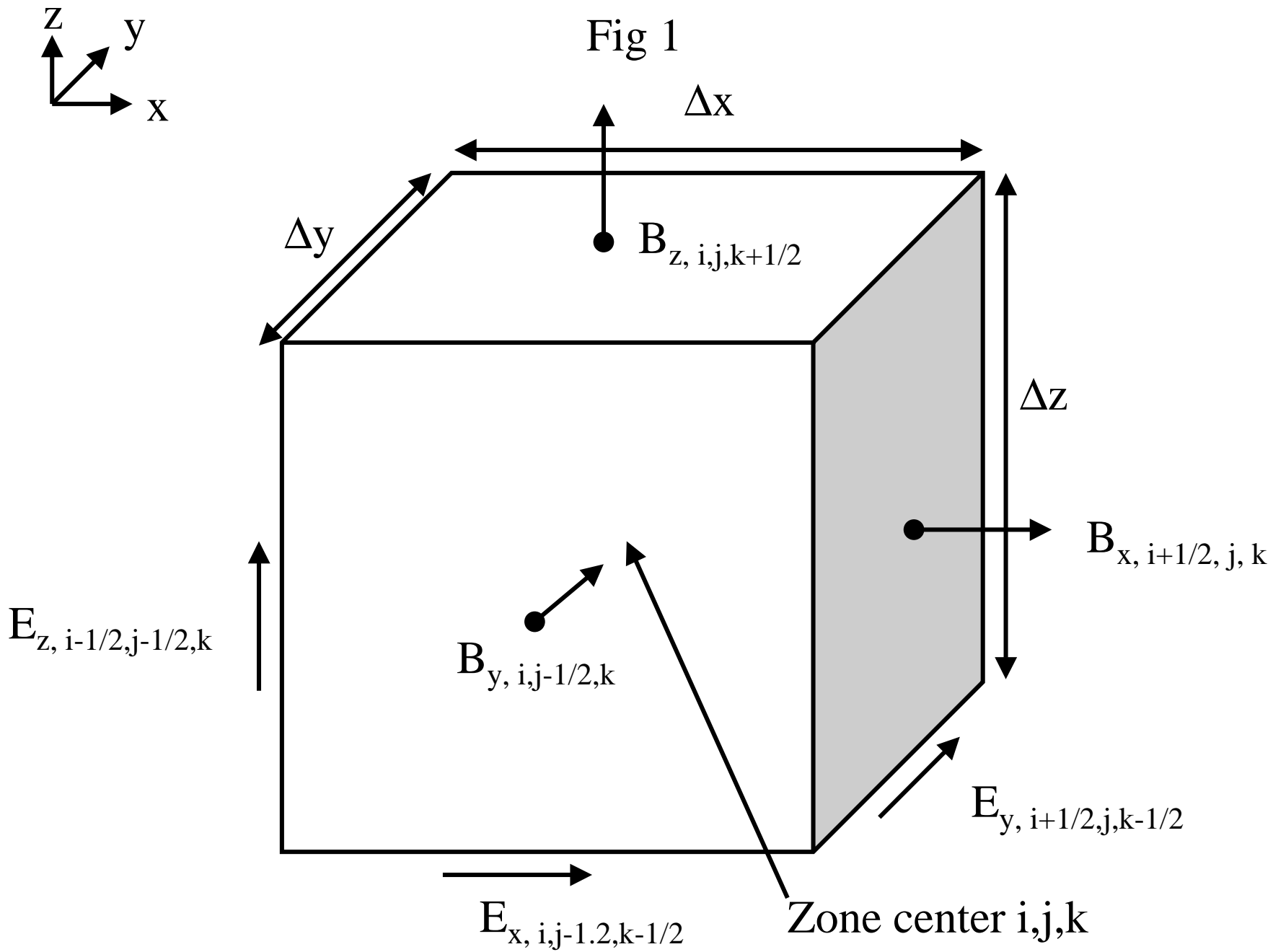

Fig 1

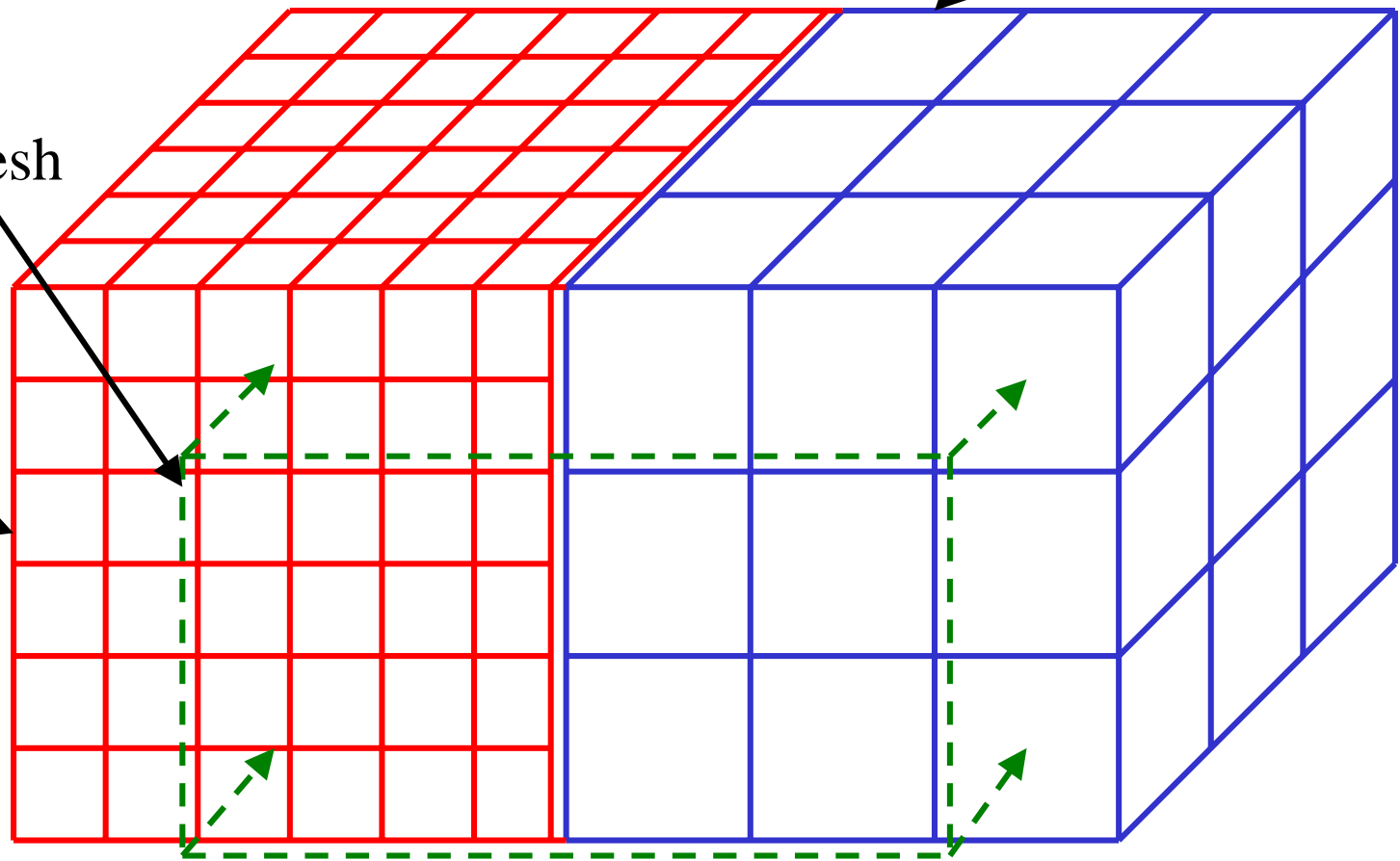

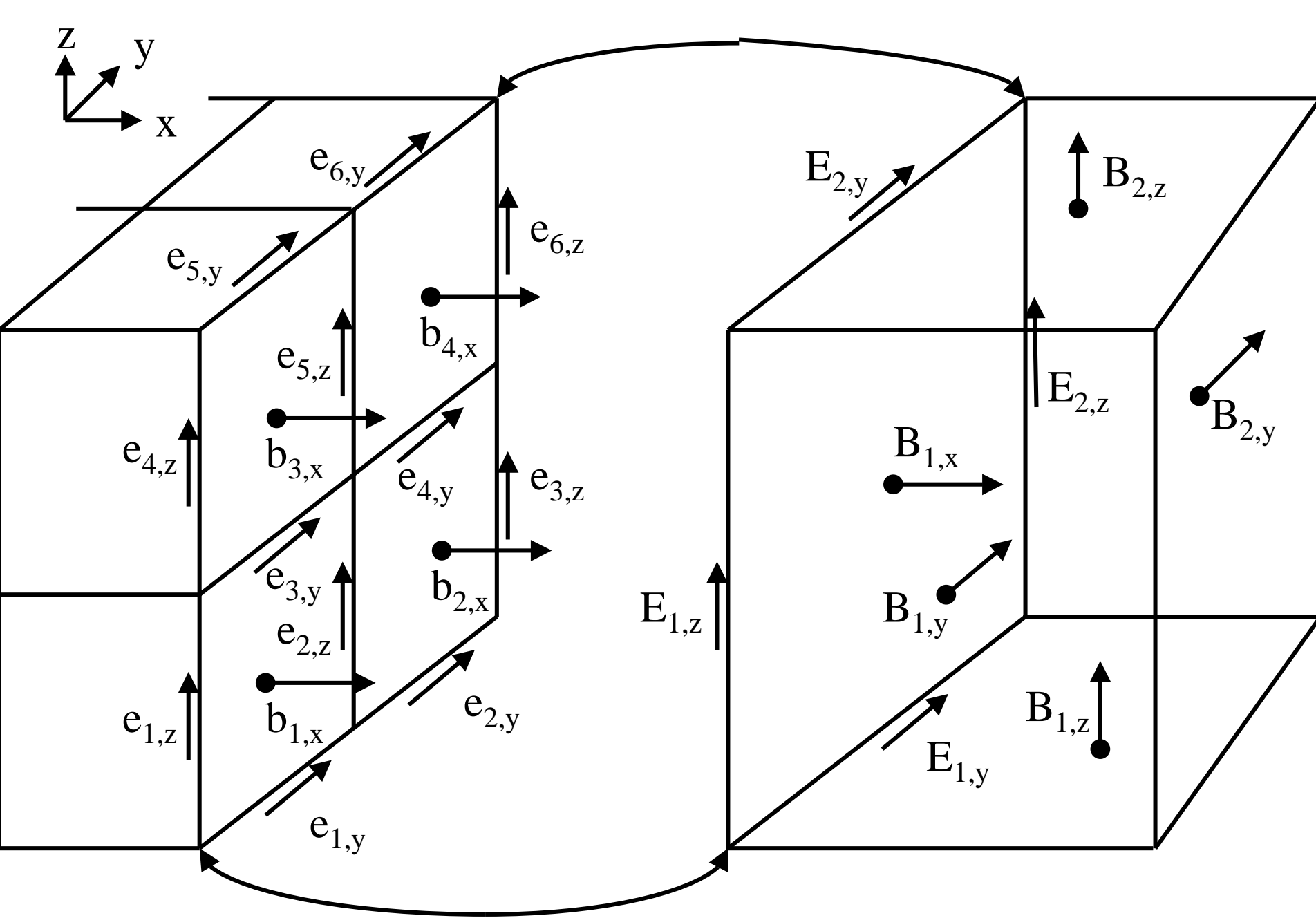

Fig 3

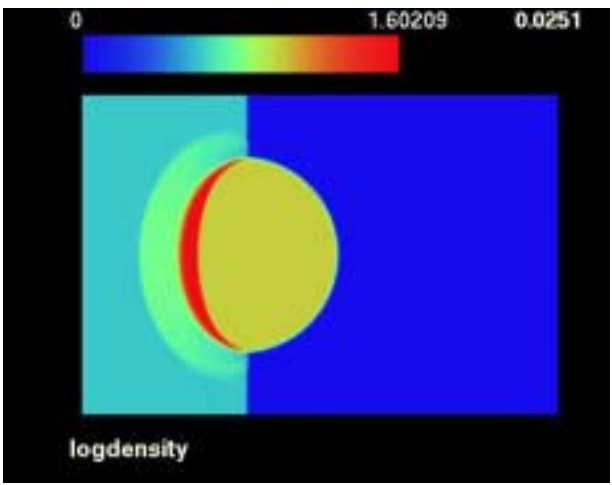

Fig 4a

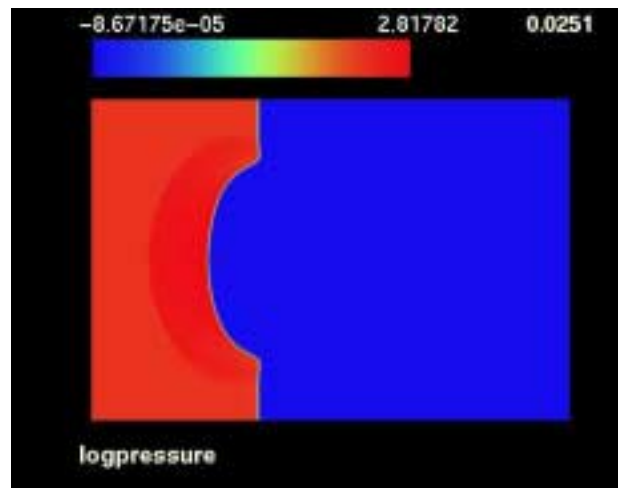

Fig 4b

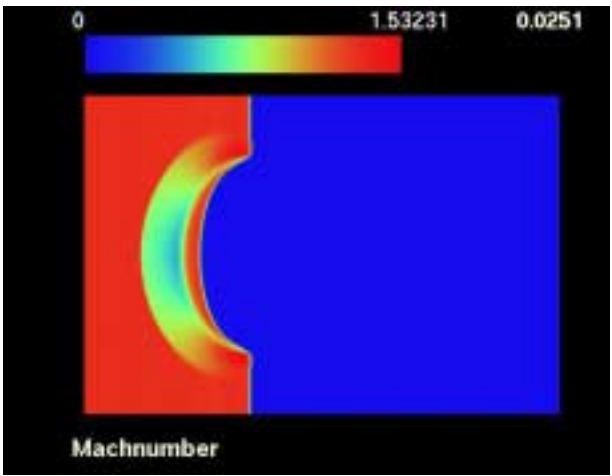

Fig 4c

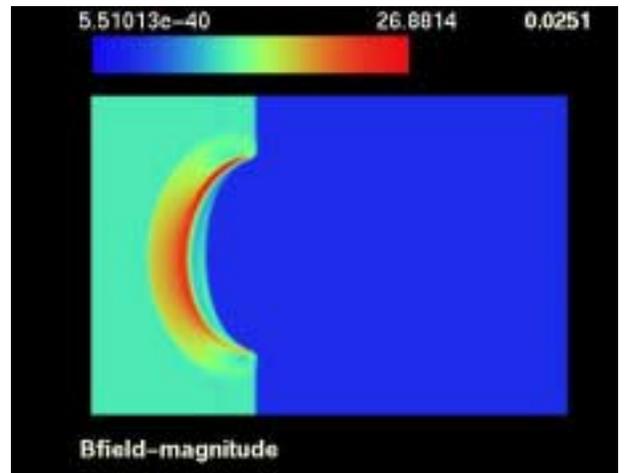

Fig 4d

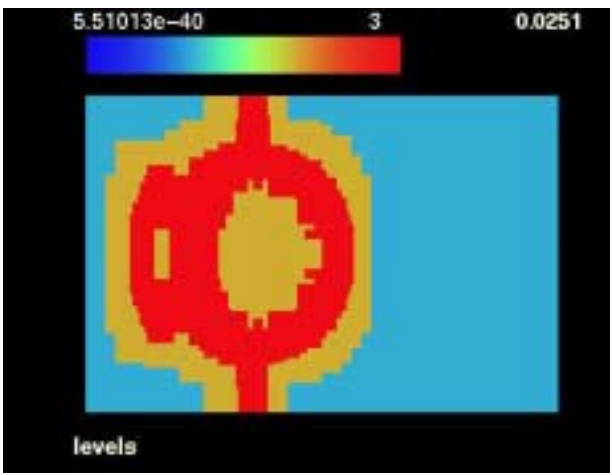

Fig 4e

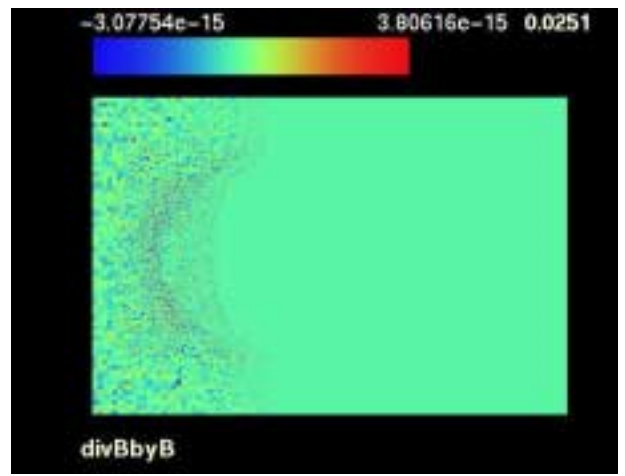

Fig 4f

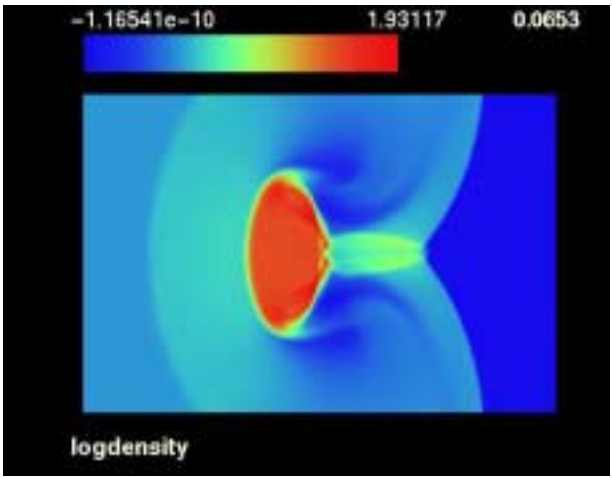

Fig 5a

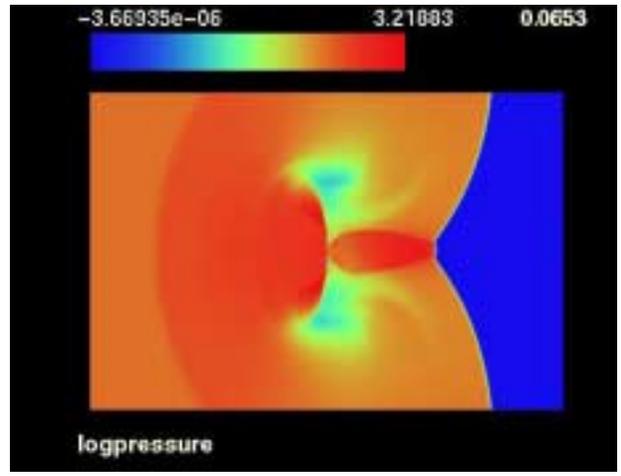

Fig 5b

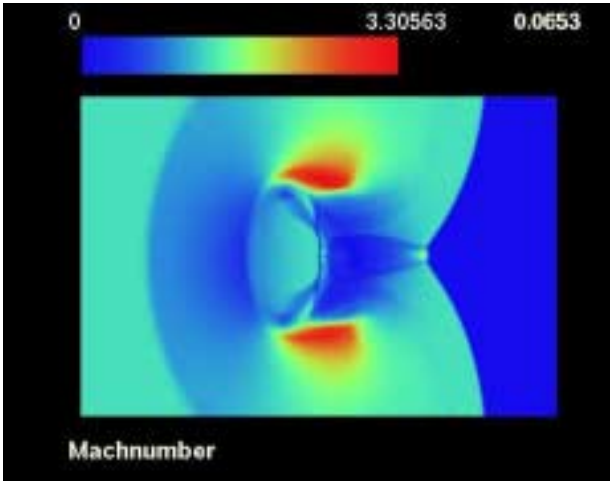

Fig 5c

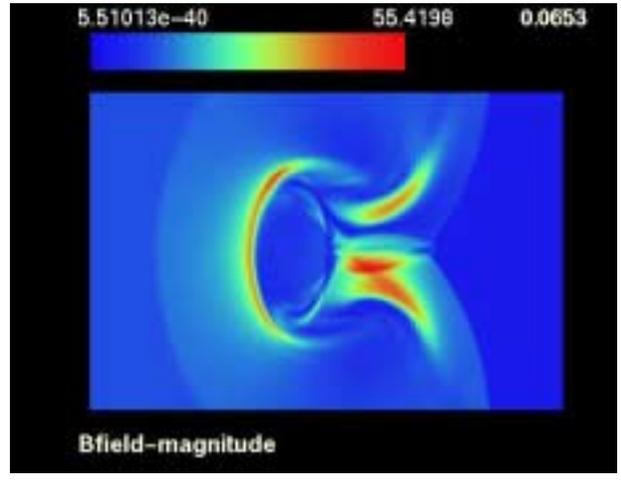

Fig 5d

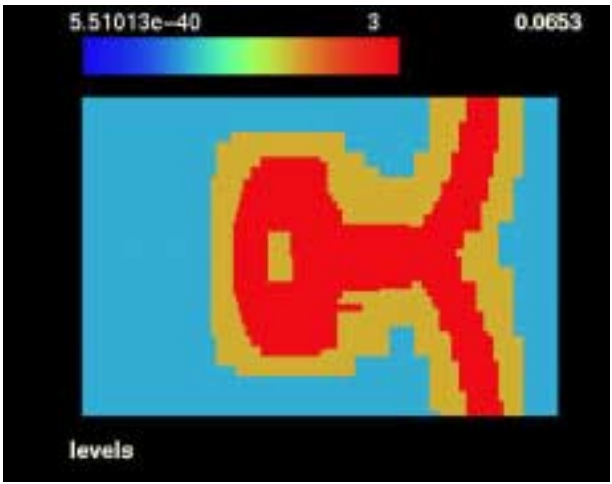

Fig 5e

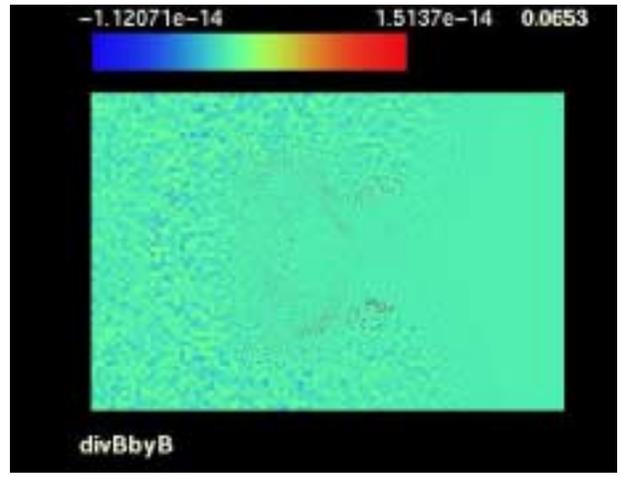

Fig 5f

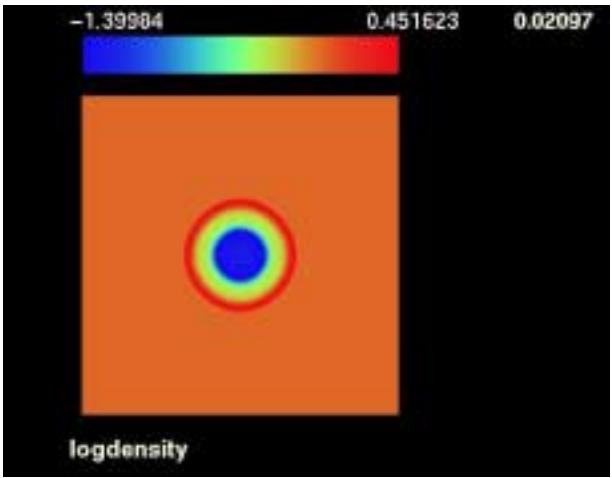

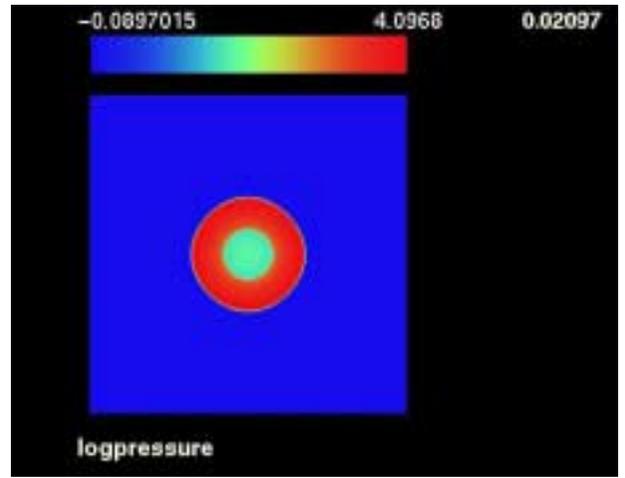

Fig 6a                                    Fig 6b

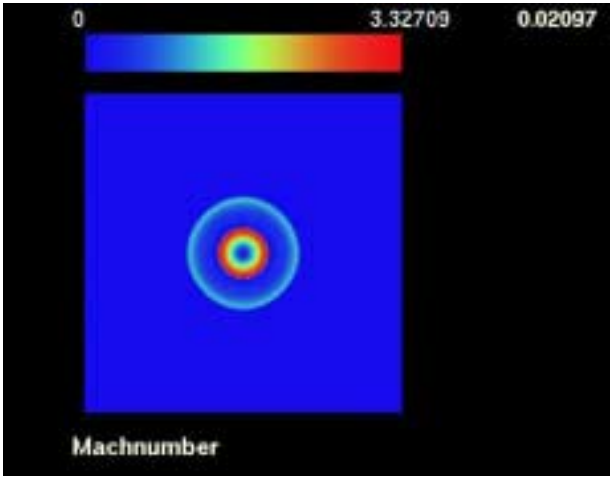

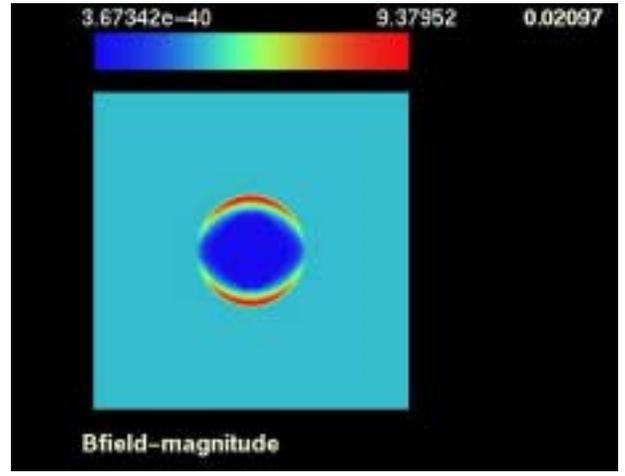

Fig 6c                                    Fig 6d

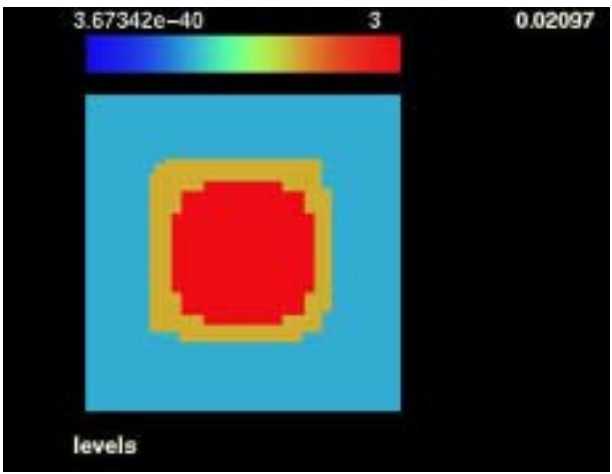

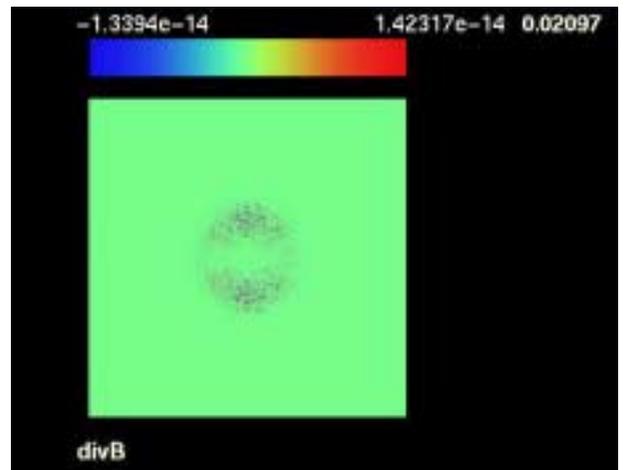

Fig 6e                                    Fig 6f

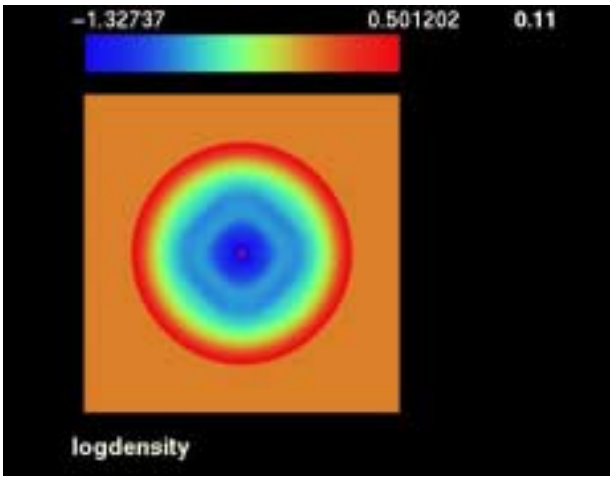

Fig 7a

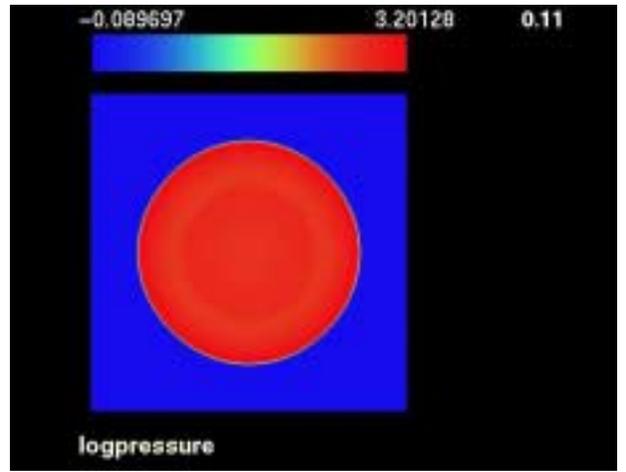

Fig 7b

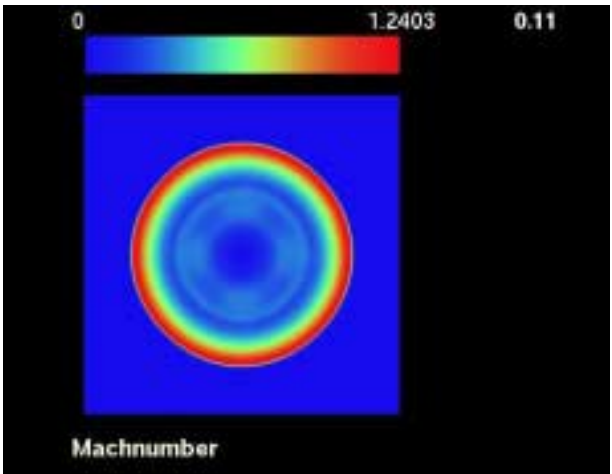

Fig 7c

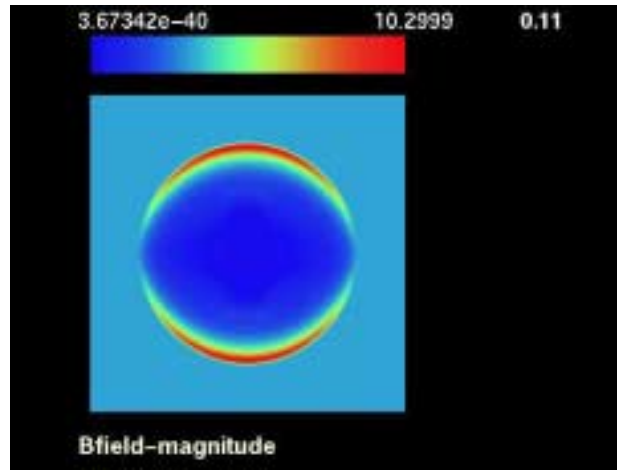

Fig 7d

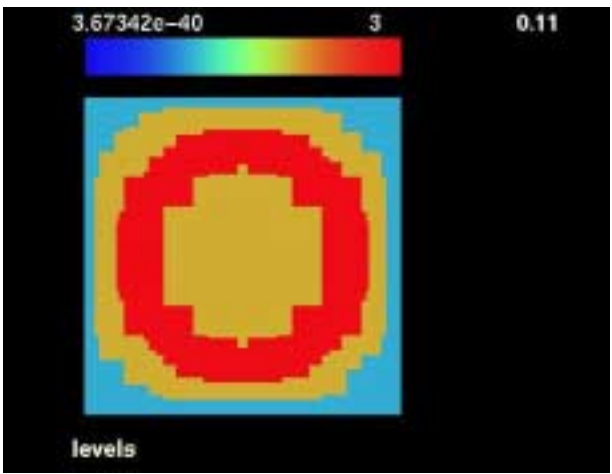

Fig 7e

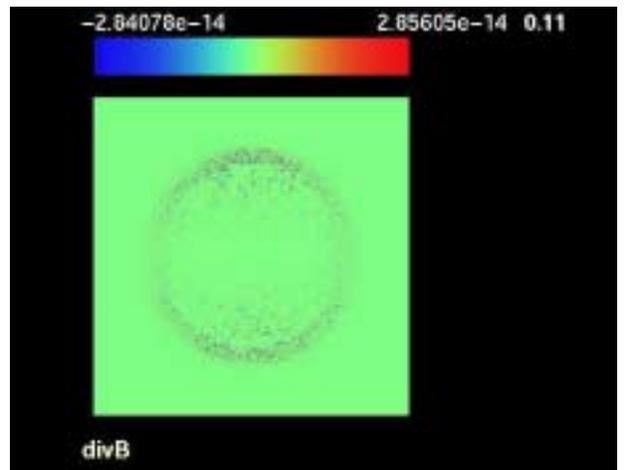

Fig 7f